\documentstyle[11pt]{article}
\begin{document}
\setlength{\oddsidemargin}{0.9in}
\setlength{\evensidemargin}{0.9in}
\baselineskip 0.50cm
\title{
Rotating embedded black holes: Entropy and Hawking's radiation}

\author {Ng. Ibohal\\
Department of Mathematics,\\
Manipur University, Imphal 795003, Manipur, INDIA.\\
E-mail: (i) ngibohal@iucaa.ernet.in (ii) ngibohal@rediffmail.com}
\date {December 27, 2004}

\maketitle

\begin{abstract}
In this paper, by applying Newman-Janis algorithm to a spherically
symmetric `seed' metric, we present general rotating metrics in
terms of Newman-Penrose (NP) quantities involving Wang-Wu
functions. From these NP quantities we present a class of
rotating solutions including (i) Vaidya-Bonnor, (ii)
Kerr-Newman-Vaidya, (iii) de Sitter, (iv) Kerr-Newman-Vaidya-de
Sitter and (v) Kerr-Newman-monopole. The rotating
Kerr-Newman-Vaidya solution represents a black hole that the
Kerr-Newman black hole is embedded into the rotating Vaidya
radiating universe. In the case of Kerr-Newman-Vaidya-de Sitter
solution, one can describe it as the Kerr-Newman black hole is
embedded into the rotating Vaidya-de Sitter universe, and
similarly, Kerr-Newman-monopole. We have also discussed the
physical properties by observing the energy momentum tensors of
these solutions. These embedded solutions can be expressed in
Kerr-Schild forms describing the extensions of Glass and Krisch
superposition, which is further the extension of Xanthopoulos
superposition. It is shown that, by considering the charge to be 
a function of radial coordinate, the Hawking's continuous radiation
of black holes can be expressed in classical spacetime metrics 
for these embedded black holes. It is also found that the 
electrical radiation will continue to form `instantaneous' 
charged black holes and creating embedded negative mass naked 
singularities describing the possible life style of radiating 
embedded black holes during their contineous radiation processes.
The surface gravity, entropy and angular velocity, which are 
important parameters of a horizon, are also 
presented for each of the embedded black holes.
\end{abstract}

\begin{twocolumn}

\setcounter{equation}{0}
\renewcommand{\theequation}{1.\arabic{equation}}

\begin{center}
{\bf 1. INTRODUCTION}
\end{center}
\setcounter{equation}{0}
\renewcommand{\theequation}{1.\arabic{equation}}

In the theory of black hole expressed in spherical polar
coordinates, there is a singularity at the origin $r=0$; whereas
at infinity $r=\infty$, the metric approaches the Minkowski flat
space. The event horizons of a black hole are expressed by this
coordinate. Thus, the nature of black holes depends on the radial
coordinate $r$. In an earlier paper [1] it is shown that Hawking's
radiation [2] can be expressed in classical spacetime metrics, by
considering the charge $e$ to be function of the radial coordinate
$r$ of non-rotating Reissner-Nordstrom as well as rotating
Kerr-Newman black holes. The variable-charge $e(r)$ with respect
to the coordinate $r$ is followed from Boulware's suggestion [3]
that {\it the stress-energy tensor may be used to calculate the
change in the mass due to the radiation}. According to Boulware's
suggestion, the energy momentum tensor of a particular space can
be used to calculate the change in the mass in order to
incorporate the Hawking's radiation effects in classical spacetime
metrics. This idea suggests to consider the stress-energy tensor
of electromagnetic field of different forms or functions from
those of Reissner-Nordstrom, as well as Kerr-Newman, black holes
as these two black holes do not seem to have any direct Hawking's
radiation effects. Thus, a variable charge in the field equations
will have the different function of the energy momentum tensor of
the charged black hole. {\sl Such a variable charge $e$ with
respect to the coordinate $r$ in Einstein's equations is referred
to as an electrical radiation ({\rm or} Hawking's electrical
radiation) of the black hole}. So, for every electrical radiation
we consider the charge $e$ to be a function of $r$ in solving the
Einstein-Maxwell field equations and we have shown {\sl
mathematically} how the electrical radiation induces to produce
the changes of the mass of {\sl variable}-charged black holes. One
may incorporate the idea of losing (or changing) mass at the rate
as the electrical energy is radiated from the charged black hole.
In fact, the change in the mass of a charged black hole takes
place due to the vanishing of Ricci scalar of the electromagnetic
field. Every electrical radiation $e(r)$ of the black holes leads
to a reduction in its mass by some quantity. If we consider such
electrical radiation taking place continuously for a long time,
then a continuous reduction of the mass will take place in the
black hole body, and the original mass of the black hole will
evaporate completely. At that stage the complete evaporation of
the mass will lead the gravity of the object depending only on the
electromagnetic field, and not on the mass. We refer to such an
object with zero mass as an `instantaneous' naked singularity - a
naked singularity that exists for an instant and then continues
its electrical radiation to create negative mass [1]. So this
naked singularity is different from the one mentioned in
Steinmular {\it et al} [4], Tipler {\it et al} [5] in the sense
that an `instantaneous' naked singularity, discussed in [4,5]
exists only for an instant and then disappears.

It is also noted that the time taken between two consecutive
radiations is supposed to be so short that one may not physically
realize how quickly radiations take place. Thus, it seems natural
to expect the existence of an `instantaneous' naked singularity
with zero mass only for an instant before continuing its next
radiation to create a negative mass naked singularity. This
suggests that it may also be possible in the common theory of
black holes that, as a black hole is invisible in nature, one may
not know whether, in the universe, a particular black hole has
mass or not, but electrical radiation may be detected on the black
hole surface. Immediately after the complete evaporation of the
mass, if one continues to radiate the remaining remnant, there may
be a formation of a new mass. If one repeats the electrical
radiation further, the new mass might increase gradually and then
the spacetime geometry will represent the negative mass naked
singularity. The classical spacetime metrics, for both stationary
rotating and non-rotating, which represent the negative mass naked
singularities have given in [1].

The aim of this paper is to give examples of rotating charged
solutions of Einstein's field equations from the general metric
for studying the Hawking's electrical radiation effects. The
solutions derived here describe rotating embedded black holes
i.e., the Kerr-Newman black hole is embedded into (i) the rotating
Vaidya null radiating space, (ii) the rotating Vaidya-de Sitter
cosmological universe and (iii) the rotating monopole space to
generate the Kerr-Newman-Vaidya, the Kerr-Newman-Vaidya-de Sitter
and the Kerr-Newman-monopole black holes respectively. The
definitions of these embedded black holes are in agreement with
the one defined by Cai et at. [6] , that {\sl when the
Scharzschild black hole is embedded into the de Sitter space, one
has the Schwarzschild-de Sitter black hole}. Thus, these solutions
describe new {\sl non-stationary} rotating, Kerr-Newman-Vaidya and
Kerr-Newman-Vaidya-de Sitter, and {\sl stationary}
Kerr-Newman-monopole black holes. These embedded rotating
solutions can be expressed in Kerr-Schild forms to regard them as
the extension of Glass-Krisch superposition [7], which is further
the extension of that of Xanthopoulos [8]. These Kerr-Schild
ansatze show that these embedded black holes are solutions of
Einstein's field equations. It is also noted that such generation
of embedded solutions in {\sl non-rotating} cases can be seen in
[9]. Here we try to extend the earlier results [1] based on
Hawking's radiation of {\sl non-embedded}, Reissner-Nordstrom as
well as Kerr-Newman, black holes to these {\sl embedded}
Kerr-Newman-Vaidya, Kerr-New-\\man-Vaidya-de Sitter and
Kerr-Newman-monopole black holes.

This paper is organized as follows: Section 2 deals with the
application of Newman-Janis algorithm to a spherically symmetric
`seed' metric with the function $M$ and $e$ of two variables $u,r$
and the presentation of general metric in terms of Newman-Penrose
(NP) quantities. In section 3, by using the NP quantities of the
general metric obtained in section 2, we derive a class of
rotating solutions, including Kerr-Newman-Vaidya, de Sitter and
Kerr-Newman-Vaidya-de Sitter, Kerr-Newman-monopole black holes. We
discuss the physical properties of the solutions observing
the nature of their energy momentum tensors and Weyl scalars.
We also present the surface gravity, entropy and angular velocity
for each of these embedded black holes as they are important 
parameters of a black hole. Section 4 deals with the 
Hawking's radiation on the variable-charged, Kerr-Newman-Vaidya, 
Kerr-Newman-Vaidya-de
Sitter and Kerr-Newman-monopole black holes. We present various
classical spacetime metrics affected by the change in the masses
describing the possible life style of radiating embedded black
holes in different stages during radiation process. In section 5
we conclude with a discussion of our results. The spin
coefficients, the Weyl scalars and the Ricci scalars for the
rotating metric discussed here are, in general, cited in an
appendix for future use.

Here, as in [1] it is convenient to use the phrase `{\sl change in
the mass}' rather than `{\sl loss of mass}' as there may be a
possibility of creation of mass after the exhaustion of the
original mass if one continues the same process of electrical
radiation. This will be seen later in the paper. The presentation
of this paper is essentially based on the Newman-Penrose (NP)
spin-coefficient formalism [10]. The NP quantities are calculated
through the technique developed by McIntosh and Hickman [11] in
(-2) signature.

\begin{center}
{\bf 2. NEWMAN-JANIS ALGORITHM AND GENERAL METRICS}
\end{center}
\setcounter{equation}{0}
\renewcommand{\theequation}{2.\arabic{equation}}

To begin with we consider a
spherical symmetric `seed' metric
written in the form
\begin{equation}
ds^2=e^{2\phi}\,du^2+2du\,dr-r^2(d\theta^2
+{\rm sin}^2\theta\,d\phi^2),
\end{equation}
where
\begin{eqnarray*}
e^{2\phi} = 1-2M(u,r)/r+e^2(u,r)/r^2
\end{eqnarray*}
and the coordinates chosen are $\{x^1,x^2,x^3,x^4\}$
\\=$\{u,r,\theta,\phi\}$. The u-coordinate is related to the
retarded time in flat space-time. So u-constant surfaces are null
cones open to the future. The r-constant is null coordinate. The
$\theta$ and $\phi$ are usual angle coordinates. The retarded time
coordinates are used to evaluate the radiating (or outgoing)
energy momentum tensor around the astronomical body [10]. Here $M$
and $e$ are the metric functions of the retarded time coordinate
$u$ and the radial coordinate $r$. Initially, when $M$, $e$ are
constant, this metric provides the non-rotating Reissner-Nordstrom
solution and also when both $M$, $e$ are functions of $u$, it
becomes the non-rotating Vaidya-Bonnor solution [13].

Now we apply Newman-Janis (NJ) algorithm [14] which is a complex
coordinate transformation,
\begin{eqnarray}
r &=& r'-i\,a\,{\rm cos}\theta,\: u = u'+i\,a\,{\rm
cos}\theta, \nonumber \\
\theta &=& \theta',\:
\phi = \phi'.
\end{eqnarray}
to make the metric (2.1) rotation. This complex transformation can
be done only when $r'$ and $u'$ are considered to be real. All the
primes are being dropped for convenience of notation. The
application of NJ algorithm is also employed by various authors
[15,16,17,18] to different `seed' metrics. Then the transformed
metric takes the following form
\begin{eqnarray}
d s^2&=&e^{2\phi}\,du^2+2du\,dr+2a\,{\rm
sin}^2\theta(1-e^{2\phi})\,du\,d\phi \cr &&-2a\,{\rm
sin}^2\theta\,dr\,d\phi -R^2d\theta^2\cr &&-\{R^2-a^2\,{\rm
sin}^2\theta\,(e^{2\phi} -2)\}\,{\rm sin}^2\theta\,d\phi^2,
\end{eqnarray}
where
\begin{equation}
e^{2\phi} = 1-{2rM(u,r,\theta)\over
R^2}+{e^2(u,r,\theta)\over R^2}.
\end{equation}
and
$R^2 = r^2 + a^2{\rm cos}^2\theta$.
It is noted that after the complex coordinate transformation
(2.2), $M$ and $e$ should be arbitrary functions
of three variables $u, r,\theta$. Then
the covariant complex null tetrad vectors take the forms
\begin{eqnarray}
&\ell_a =& \delta^1_a -a\,{\rm sin}^2\theta\,\delta^4_a, \\
&n_a =&{1\over 2}\,H\,\delta^1_a+ \delta^2_a -{1\over
2}\,H\,a\,{\rm sin}^2\theta\,\delta^4_a, \cr &m_a =&-{1\over\surd
2R}\,\{-ia\,{\rm
sin}\theta\,\delta^1_a+R^2\,\delta^3_a  \nonumber \\
&&+i(r^2+a^2)\,{\rm sin}\,\theta\,\delta^4_a\}.
\end{eqnarray}
The null vectors $\ell_a$ and $n_a$ are real, and $m_a$
is complex null vector with the normalization condition
$\ell^a\,n_a$=$-m^a\,\overline
m_a = 1$ and $R=r+ia\,{\rm cos}\theta$. Here,
\begin{equation}
H(u,r,\theta) = R^{-2}\{r^2-2rM(u,r,\theta)+a^2
+e^2(u,r,\theta)\}.
\end{equation}

The NP quantities ({\it i.e.} the NP spin coefficients, the Ricci
and Weyl scalars cited in appendix below) have been calculated
with the arbitrary metric functions $M$ and $e$ of three variables
of the metric (2.3). From the NP spin coefficients, it is found
that the rotating spherically symmetric metric (2.3)  possesses,
in general, a geodesic $(\kappa=\epsilon=0)$, shear free
$(\sigma=0)$, expanding $(\hat{\theta} \neq 0)$ and non-zero twist
$(\hat{\omega}^{2} \neq 0)$  null vector $\ell_a$ [19] where
\begin{eqnarray}
&&\hat{\theta}\equiv -{1\over2}(\rho + \overline\rho) ={r\over
R^2}, \\ &&\hat{\omega}^{2}\equiv-{1\over4}(\rho -
\overline\rho)^2 =-{{a^2\cos^2\theta}\over {R^2\,R^2}}.
\end{eqnarray}

From the Einstein's equations,
\begin{equation}
G_{ab}\equiv R_{ab} - {1\over 2}\,R\,g_{ab} = - K\,T_{ab},
\end{equation}
we obtain the energy momentum tensor (EMT) for the  metric (2.3)
as follows:
\begin{eqnarray}
T_{ab} &=&\mu^*\,\ell_a\,\ell_b+
2\,\rho^*\,\ell_{(a}\,n_{b)}
+2\,p\,m_{(a}\overline m_{b)} \cr
& & + 2\,\omega\,\ell_{(a}\,\overline m_{b)} +
2\,\overline\omega\,\ell_{(a}\,m_{b)},
\end{eqnarray}
where $\mu^*$, $\rho^*$, $p$ and $\omega$ are related  to the
null density, the matter density,
the pressure $p$ as well as the rotation function
respectively of the rotating spherically symmetric objects
and are given in terms of Ricci scalars as follows:
\begin{eqnarray}
&&K\,\rho^* =  2\,\phi_{11} + 6\,\Lambda,\;\; K\,p = 2\,\phi_{11}
- 6\,\Lambda,\cr &&K\,\mu^* = 2\,\phi_{22},\;\;  K\,\omega = -
2\,\phi_{12},
\end{eqnarray}
where $\phi_{11}$, $\phi_{12}$, $\phi_{22}$, $\Lambda$ are the
non-vanishing Ricci scalars given in appendix (A3) for the metric
(2.3). Then we have
\begin{eqnarray}
&\mu^* =&-{1\over K\,R^2\,R^2}\,\Big\{ 2r(r\,M_{,u}-ee_{,u})\cr
&&-{\rm cot}\theta(rM_{,\theta}-ee_{,\theta}) + a^2{\rm
sin}^2\theta(rM_{,u}-ee_{,u})_{,u}\cr &&-(rM_{,\theta}
-ee_{,\theta})_{,\theta}\Big\}, \cr &\rho^*=&{1\over
K\,R^2\,R^2}\Big\{e^2+2\,r(r\,M_{,r}
-e\,e_{,r})\Big\}, \\
&p=&{1\over K\,R^2\,R^2}\Big\{e^2+2\,r(r\,M_{,r}
-e\,e_{,r}) \nonumber \\
&&-R^2(2M_{,r}+r\,M_{,r\,r}-e^2_{,r}-e\,e_{,r\,r})\Big\}, \cr
&\omega =&-{1\over \surd 2\,K\,R^2\,R^2}\,\Big[i\,a\,{\rm
sin}\theta \Big\{(R\,M_{,u}-2\,e\,e_{,u}) \cr
&&-(r\,M_{,r}-e\,e_{,r})_{,u}\overline R\Big\} \cr
&&+\Big\{(R\,M_{,\theta}-2\,e\,e_{,\theta})-(
r\,M_{,r}-e\,e_{,r})_{,\theta}\,\overline R\Big\}\Big]. \nonumber
\end{eqnarray}
Here the Ricci scalar $\Lambda\equiv (1/24)g^{ab}\,R_{ab}$ for the
metric (2.3) is
\begin{eqnarray}
\Lambda={1\over{12\,R^2}}\,\Big(2M_{,r}+
r\,M_{,r\,r}-e^2_{,r}-ee_{,r\,r}\Big).
\end{eqnarray}
It is observed that the expression of
$\Lambda$ does not involve any derivative of $M$ and $e$ with respect to
$u$ and $\theta$, though $M$ and $e$ are functions
of three variables $u, r,\theta$.

The above EMT (2.11) can be written as  $T^{(\rm n)}_{ab}$ and
$T^{(\rm m)}_{ab}$ to represent two fluid systems {\it i.e.}
rotating null fluid $T^{(\rm n)}_{ab}$ and  rotating
matter $T^{(\rm m)}_{ab}$. Then we have
\begin{eqnarray}
T^{(\rm n)}_{ab}&=& \mu^*\ell_a\ell_b + \omega\ell_{(a}\overline
m_{b)}+\overline\omega\ell_{(a}\,m_{b)}, \\
T^{(\rm m)}_{ab}&=&2\,(\rho^*+p)\ell_{(a}\,n_{b)} - p\,g_{ab} \cr
&& + \omega\,\ell_{(a}\,\overline m_{b)} +
\overline\omega\,\ell_{(a}\,m_{b)},
\end{eqnarray}
where $\overline\omega$ is the complex conjugate of $\omega$. The
appearance of non-vanishing $\omega$ in $T^{(\rm n)}_{ab}$ and
$T^{(\rm m)}_{ab}$ shows the rotating fluid systems in spherically
symmetric spacetime geometry. When we set $\omega = 0$ initially,
these EMTs may be similar to those introduced by Husain [20] and
Glass and Krisch [7] with the arbitrary mass $M(u,r)$ and charge
$e(u,r)$ for the non-rotating objects. From the appendix (A3) we
find that $\phi_{00} = 0$. So the vanishing of $\phi_{00}$
suggests the possibility that the metric (2.3) does not possess a
perfect fluid, whose energy-momentum tensor is
$T_{ab}=(\rho^*+p)u_au_b-p\,g_{ab}$ with
$\phi_{00}=2\phi_{11}=\phi_{22}=-K(\rho^*+p)/4$,
$\Lambda=K(3p-\rho^*)/24$ with a time-like vector $u^a$.

\begin{center}
{\bf 3. ROTATING SOLUTIONS}
\end{center}
\setcounter{equation}{0}
\renewcommand{\theequation}{3.\arabic{equation}}

In this section, from the general solutions presented in appendix
we shall present a class of rotating solutions, namely (a)
Vaidya-Bonnor, (b) de Sitter, (c) Kerr-Newman-Vaidya, (d) 
Kerr-Newman-Vaidya-de Sitter,
which are to be discussed in this paper. The rotating
Vaidya-Bonnor solution descri-\\bes a non-stationary spherically
symmetric solution of Einstein's field equations. The rotating de
Sitter solution is a Petrov type $D$ solution. The rotating
Kerr-Newman-Vaidya solution represents a non-stationary
black-hole, describing Kerr-Newman black hole embedded into the
rotating Vaidya null radiating universe. Again the rotating
Kerr-Newman-Vaidya-de Sitter solution describes the non-stationary
Kerr-Newman-Vaidya black hole embedded into the rotating de
Sitter cosmological universe.

\vspace*{0.15in}

{\it (i) Rotating charged Vaidya-Bonnor solution}: \\
$M=M(u)$, $a\neq 0$, $e=e(u)$:

\vspace{.15in}

Once the restrictions on $M$ and $e$ are considered to be
functions of $u$ only, the quantities (2.13) become quite simple.
In this case the energy momentum tensor (2.11) takes
\begin{eqnarray}
T_{ab}&=& \mu^*\,\ell_a\,\ell_b + 2\,\rho^*\,\{\ell_{(a}\,n_{b)}
+m_{(a}\overline m_{b)}\} \cr &&+ 2\,\omega\,\ell_{(a}\,\overline
m_{b)} + 2\,\overline\omega\,\ell_{(a}\,m_{b)},
\end{eqnarray}
with
\begin{eqnarray}
\mu^* &=&-{1\over K\,R^2\,R^2}\Big\{2\,r\,(r\,M_{,u}-e\,e_{,u})
\cr &&+a^2{\rm sin}^2\theta\,(r\,M_{,u}-e\,e_{,u})_{,u}\Big\}, \cr
\rho^*
&=& p = {e^2(u)\over K\,R^2\,R^2}, \\
\omega &=&{-i\,a\,{\rm sin}\,\theta\,\over{\surd
2\,K\,R^2\,R^2}}\,\Big\{R\,M_{,u}-2e\,e_{,u}\Big\}, \nonumber
\end{eqnarray}
and the Weyl scalars are
\begin{eqnarray}
\psi_2&=&{1\over\overline R\,\overline R\,R^2}\Big(e^2 -R\,M\Big),
\cr \psi_3 &=&{-i\,a\,{\rm sin}\theta\over 2\surd 2\overline
R\,\overline R\,R^2}\Big\{4\,(rM_{,u}-e\,e_{,u})+\overline
R\,M_{,u}\Big\}, \cr \psi_4 &=&{{a^2\,{\rm sin}^2\theta}\over
2\overline R\,\overline R\,R^2\,R^2}\,\Big\{
R^2\,(r\,M_{,u}-e\,e_{,u})_{,u} \cr
&&-2r\,(r\,M_{,u}-e\,e_{,u})\Big\}.
\end{eqnarray}
The line element will take the form
\begin{eqnarray}
ds^2&=&[1-\{2rM(u)-e^2(u)\}R^{-2}]\,du^2+2du\,dr \cr
&&+2aR^{-2}\{2rM(u)-e^2(u)\}{\rm sin}^2\theta\,du\,d\phi \cr
&&-2a\,{\rm sin}^2\theta\,dr\,d\phi -R^2d\theta^2 -\{(r^2+a^2)^2
\cr &&- \Delta^*a^2\,{\rm sin}^2\theta\}R^{-2}{\rm
sin}^2\theta\,d\phi^2,
\end{eqnarray}
where $\Delta^*=r^2-2rM(u)+a^2+e^2(u)$. This solution describes a
black hole when $M(u)>a^2+e^2(u)$ and has $r_{\pm}=M(u)^*\pm \surd
{\{M^2(u)-a^2-e^2(u)\}}$ as the roots of the equation
$\Delta^*=0$. So the rotating Vaidya-Bonnor solution has an {\it
external event horizon} at $r=r_{+}$ and an {\it internal Cauchy
horizon} at $r=r_{-}$. The non-stationary limit surface $g_{uu}>0$
of the rotating black hole {\it i.e.} $r\equiv
r_e(u,\theta)=M(u)+\surd {\{M^2(u)-a^2{\rm cos}^2\theta-e^2(u)\}}$
does not coincide with the event horizon at $r_+$, thereby
producing the ergosphere. The surface gravity of the event horizon
at $r=r_+$ is
\begin{eqnarray*}
{\cal K}=-\frac{1}{r_+R^2}\Big[r_+\sqrt{\{f(u)^2-a^2+e^2(u)
\}}+\frac{e^2(u)}{2}\Big],
\end{eqnarray*}
and the entropy of the horizon is  
\begin{eqnarray*}
{\cal S}=2\pi\,f(u)\Big[f(u)+\sqrt{f(u)^2
-a^2-e^2(u)}\,\Big]-\frac{e^2}{4}.
\end{eqnarray*}
The angular velocity of the horizon is given by 
\begin{eqnarray*}
\Omega_{\rm H}=\frac{a\{2rM(u)-e^2(u)\}}{(r^2+a^2)^2}\Big|_{r=r_{+}}.
\end{eqnarray*}
We have seen the direct involvement of the rotation parameter $a$ 
in both the expressions of surface gravity and the angular velocity,
showing the different structure of rotating black hole. When $a=0$,
the angular velocity will also vanish for the horizon.

From the rotating Vaidya-Bonnor metric,  we can clearly  recover 
the following solutions: (a) rotating Vaidya metric when $e(u)=0$, 
(b) rotating charged Vaidya solution when $e(u)$ becomes constant, 
(c) the rotating Kerr-Newman solution when $M(u) = e(u)$ = 
constant and (d) well-known non-rotating Vaidya-Bonnor metric [13]
when $a=0$. It is also noted that when $e=a=0$, the null density
of Vaidya radiating fluid takes the form $\mu^*
=-{2\,M_{,u}/K\,r^2}$. The non-rotating Vaidya null radiating
metric is of type $D$ in the Petrov classification of spacetime,
whose one of the repeated principal null vectors, $\ell_a$ is a
geodesic, shear free, non-rotating with non-zero expansion [21],
while the rotating one is of algebraically special with a null
vector $\ell_a$ (2.5), which is geodesic, shear free, expanding as
well as non-zero twist. The rotating Vaidya-Bonnor metric (3.4)
can be expressed in Kerr-Schild ansatz on the rotating Vaidya null
radiating background as
\begin{eqnarray*}
g_{ab}^{\rm VB}=g_{ab}^{\rm V} +2Q(u,r,\theta)\ell_a\ell_b
\end{eqnarray*}
with $Q(u,r,\theta) = \{e^2(u)/2R^2\}$, indicating the existence 
of electromagnetic field on the rotating Vaidya spacetime geometry.

Carmeli and Kaye [22] have also obtained the rotating Vaidya
metric (a) above and discussed under the name of a variable-mass
Kerr solution. Herrera and Martinez [23] and Herrera {\it et al} 
[24] have discussed the
physical interpretation of the solution of Carmeli and Kaye.
Similarly, during the application of Newman-Janis algorithm to the
{\it non-rotating} Vaidya-Bonnor `seed' solution with $M(u)$ and
$e(u)$, Jing and Wang [25] kept the functions $M(u)$ and $e(u)$
unchanged and studied the nature of the transformed metric with
the consequent NP quantities.

\vspace{.15in}

{\it (ii) Rotating solutions with $M=M(u,r)$, $a \neq 0$,
$e(u,r,\theta)=0$}:

\vspace{.15in}

If we take $M$ to be the function of $u, r$ and
\\$e(u,r,\theta)=0$ in (2.13), the energy momentum tensor will
take the form
\begin{eqnarray}
T_{ab}&=& \mu^*\,\ell_a\,\ell_b +
2\,\rho^*\,\ell_{(a}\,n_{b)}
+2\,p\,m_{(a}\overline m_{b)} \cr
&&+2\,\omega\,\ell_{(a}\,\overline m_{b)}
+2\,\overline\omega\,\ell_{(a}\,m_{b)},
\end{eqnarray}
with the following quantities
\begin{eqnarray}
&&\mu^* =-{1\over K\,R^2\,R^2}\Big\{2r^2M_{,u} + a^2r\,{\rm
sin}^2\theta\,M_{,uu}\Big\}, \cr &&\rho^* =  {2\,r^2\over
K\,R^2\,R^2}\,M_{,r}, \cr &&p = -{1\over K}\,\Big\{{2\,a^2\,{\rm
cos}^2\theta \over
R^2\,R^2}\,M_{,r}+{r\over R^2}\,M_{,r\,r}\Big\}, \\
&&\omega =-{i\,a\,{\rm sin}\,\theta\over\surd
2\,K\,R^2\,R^2}\,\Big(R\,M_{,u}-r\,\overline R\,M_{,ur}\Big).
\nonumber
\end{eqnarray}
The line element will be of the form
\begin{eqnarray}
d s^2&=&\{1-2rM(u,r)R^{-2}\}\,du^2+2du\,dr \cr
&&+4arM(u,r)R^{-2}\,{\rm sin}^2 \theta\,du\,d\phi \cr &&-2a\,{\rm
sin}^2\theta\,dr\,d\phi -R^2d\theta^2 -\{(r^2+a^2)^2 \cr
&&-\Delta^*a^2\,{\rm sin}^2\theta\}\, R^{-2}{\rm
sin}^2\theta\,d\phi^2,
\end{eqnarray}
where $\Delta^*=r^2-2rM(u,r)+a^2$ and the Weyl scalars given in
(A2) become
\begin{eqnarray}
\psi_2&=&{1\over\overline R\,\overline
R\,R^2}\Big\{-R\,M+{\overline R\over 6}\,M_{,r}(4r+2\,i\,a\,{\rm
cos}\theta) \cr &&-{r\over 6}\,\overline R\,\overline
R\,M_{,rr}\Big\}, \cr \psi_3&=&-{i\,a\,{\rm sin}\theta\over{2\surd
2\overline R\,\overline R\,R^2}}\, \Big\{(4\,r+\overline
R\,)\,M_{,u}
 + r\,{\overline R}\,M_{,ur}\Big\}, \cr
\psi_4 &=&{{a^2\,r\,{\rm sin}^2\theta}\over 2\overline
R\,\overline R\,R^2\,R^2}\,\Big\{R^2\,M_{,uu}-2\,r\,M_{,u}\Big\}.
\end{eqnarray}

Wang and Wu [9] have expanded the metric function $M(u,r)$ for the
non-rotating solution $(a = 0)$ in the power of $r$
\begin{equation}
M(u,r)= \sum_{n=-\infty}^{+\infty} q_n(u)\,r^n,
\end{equation}
where $q_n(u)$ are arbitrary functions of $u$. They consider
the above sum as an integral when the `spectrum' index $n$ is
continuous. Here using this expression in equations (3.6) we can
generate rotating metrics with $a\neq 0$ as
\begin{eqnarray}
\mu^* &=&-{r\over K\,R^2\,R^2}\,\sum_{n=-\infty}^{+\infty}
\Big\{2\,q_n(u)_{,u}\,r^{n+1} \cr &&+a^2{\rm
sin}^2\theta\,q_n(u)_{,uu}\,r^n\Big\}, \cr \rho^* &=&{2\,r^2\over
K\,R^2\,R^2}\,\sum_{n=-\infty}^{+\infty} n\,q_n(u)\,r^{n-1}, \\
p &=& -{1\over KR^2}\sum_{n=-\infty}^{+\infty} nq_n(u)r^{n-1} \cr
&&\times\Big\{{2a^2\,{\rm cos}^2\theta \over R^2} +(n-1)\Big\},
\cr \omega &=&{-i\,a\,{\rm sin}\,\theta\over\surd
2\,K\,R^2\,R^2}\,\sum_{n=-\infty}^{+\infty}(R-n\overline
R)\,q_n(u)_{,u}\,r^n. \nonumber
\end{eqnarray}
Clearly these solutions (3.10) will recover \\ non-rotating
Wang-Wu solutions if one sets $a=0$. Here we find that these
rotating Wang-Wu solutions include many known as well as un-known
rotating solutions of Einstein's field equations with spherical
symmetry as shown by Wang and Wu in non-rotating cases [9]. The
functions $q_n(u)$ in (3.9) and (3.10) play a great role in
generating rotating solutions. Therefore, we hereafter refer to
$q_n(u)$ as Wang-Wu functions. Here a class of rotating
solutions can be derived from these quantities (3.10) as follows.

\vspace{.15in}

{\it (iii) Kerr-Newman-Vaidya solution}

\vspace{.15in}

Wang and Wu [9] could combine the three {\sl non-rotating}
solutions, namely monopole, de-Sitter and charged Vaidya solution
to obtain a new solution which represents a {\sl non-rotating}
monopole-de Sitter-Vaidya charged solutions. In the same way, we
wish to combine the Kerr-Newman solution with the rotating Vaidya
solution obtained above in (3.4) with $e(u)=0$, if the Wang-Wu
functions $q_n(u)$ in (3.10) are chosen such that
\begin{eqnarray}
\begin{array}{cc}
q_n(u)=&\left\{\begin{array}{ll}
m+f(u),&{\rm when }\;\;n=0\\
-e^2/2, &{\rm when }\;\;n=-1\\
0, &{\rm when }\;\;n\neq 0, -1,
\end{array}\right.
\end{array}
\end{eqnarray}
where $m$ and $e$ are constants. [Note: This constant $e$ is
assumed to be different from the notation $e(u,r,\theta)$ which
has been set to zero in subsection 3(ii), and can be seen its absence 
in (3.7)]. Then, the mass function takes the form
\begin{eqnarray*}
M(u,r)=m+f(u)-e^2/2r
\end{eqnarray*}
and other quantities are
\begin{eqnarray}
&&\rho^*= p = {e^2\over K\,R^2\,R^2}, \\
&&\mu^*={-r\over K\,R^2\,R^2}\Big\{2\,r\,f(u)_{,u}+a^2{\rm
sin}^2\theta\,f(u)_{,uu}\Big\}, \cr &&\omega ={-i\,a\,{\rm
sin}\,\theta\,\over{\surd 2\,K\,\overline R\,R^2}}\,f(u)_{,u}, \\
&&\Lambda \equiv  \frac{1}{24}g^{ab}\,R_{ab}=0,\cr
&&\gamma={1\over{2\overline R\,R^2}}\,\left[\{r-m-f(u)\}\overline
R-\Delta^*\right],\nonumber
\end{eqnarray}
and $\phi_{11}$, $\phi_{12}$, $\phi_{22}$ can be obtained from
equations (3.13) with (2.12). The Weyl scalars (3.8) become
\begin{eqnarray}
&&\psi_2={1\over\overline R\,\overline R\,R^2}\Big[e^2
-R\,\{m+f(u)\}\Big], \cr &&\psi_3 ={-i\,a\,{\rm sin}\theta\over
2\surd 2\overline R\,\overline R\,R^2}\Big\{(4\,r+\overline
R)f(u)_{,u}\Big\}, \\
&&\psi_4 ={{a^2r\,{\rm sin}^2\theta}\over 2\overline R\,\overline
R\,R^2\,R^2}\,\Big\{R^2f(u)_{,uu}-2rf(u)_{,u}\Big\}. \nonumber
\end{eqnarray}
This represents a rotating Kerr-Newman-Vaidya solution with
the line element
\begin{eqnarray}
d s^2&=&[1-R^{-2}\{2r(m+f(u))-e^2\}]\,du^2 \cr &&+2du\,dr
+2aR^{-2}\{2r(m+f(u)) \cr &&-e^2\}\,{\rm sin}^2 \theta\,du\,d\phi
-2a\,{\rm sin}^2\theta\,dr\,d\phi -R^2d\theta^2 \cr
&&-\{(r^2+a^2)^2 -\Delta^*a^2\,{\rm sin}^2\theta\}\,R^{-2}{\rm
sin}^2\theta\,d\phi^2,  \nonumber\\
\end{eqnarray}
where $\Delta^*=r^2-2r\{m+f(u)\}+a^2+e^2$. Here $m$ and $e$ are
the mass and the charge of Kerr-Newman solution, $a$ is the
rotational parameter per unit mass and $f(u)$ represents the mass
function of rotating Vaidya null radiating fluid. The solution
(3.15) will describe a black hole if $m+f(u)>a^2+e^2$ with
external event horizon at $r_{+}= \{m+f(u)\} \\ +\surd
{[\{m+f(u)\}^2-a^2-e^2]}$, an internal Cauchy horizon at
$r_{-}=\{m+f(u)\} -\surd{[\{m+f(u)\}^2 -a^2-e^2]}$ and the
non-stationary limit surface $r\equiv r_e(u,\theta)= \{m+f(u)\}
+\surd {[\{m+f(u)\}^2-a^2{\rm cos}^2\theta-e^2]}$. The surface
gravity of the event horizon at $r=r_{+}$ is
\begin{eqnarray*}
{\cal K}=-\frac{1}{r_+R^2}\Big[r_+\sqrt{\Big\{\Big(m+f(u)\Big)^2-(a^2+e^2)
\Big\}}+\frac{e^2}{2}\Big].
\end{eqnarray*}
The entropy of the horizon is given by
\begin{eqnarray*}
{\cal S}&=&2\pi\{m+f(u)\}\Big[\{m+f(u)\} \cr &&
+\sqrt{\{m+f(u)\}^2
-(a^2+e^2)}\,\Big]-\frac{e^2}{4}.
\end{eqnarray*}
The angular velocity of the horizon takes the form 
\begin{eqnarray*}
\Omega_{\rm H}=\frac{a[2r\{m+f(u)\}-e^2(u)]}{(r^2+a^2)^2}\Big|_{r=r_{+}}.
\end{eqnarray*}
When we set $f(u)=0$, the metric (3.15) recovers the usual
Kerr-Newman black hole, and if $m=0$, then it is the `rotating'
charged Vaidya null radiating black hole (3.4).

In this rotating solution, the Vaidya null fluid is
interacting with the non-null electromagnetic field whose
Maxwell scalar $\phi_1$ can be obtained from (3.12). Thus, we could
write the total energy momentum tensor (EMT) for the
rotating solution (3.15) as follows:
\begin{eqnarray}
T_{ab} = T^{(\rm n)}_{ab} +T^{(\rm E)}_{ab},
\end{eqnarray}
where the EMTs for the rotating null fluid as well as
that of the electromagnetic field are given
respectively as
\begin{eqnarray}
&T^{(\rm n)}_{ab}=& \mu^*\ell_a\ell_b + 2\omega\ell_{(a}\overline
m_{b)}+2\overline\omega\ell_{(a}m_{b)} \\
&T^{(\rm E)}_{ab}=&2\,\rho^*\,\{\ell_{(a}\,n_{b)}
 + m_{(a}\,\overline m_{b)}\}.
\end{eqnarray}
The appearance of non-vanishing $\omega$ shows the null
fluid is rotating as the expression of $\omega$
(3.13) involves the rotating parameter $a$ coupling with
$\partial f(u)/\partial u$, both non-zero quantities
for a rotating Vaidya null radiating universe.

This rotating Kerr-Newman-Vaidya metric (3.15) can be
expressed in Kerr-Schild form
on the Kerr-Newman background as
\begin{equation}
g_{ab}^{\rm KNV}=g_{ab}^{\rm KN} +2Q(u,r,\theta)\ell_a\ell_b
\end{equation}
where
\begin{equation}
Q(u,r,\theta) =-rf(u)R^{-2},
\end{equation}
and the vector $\ell_a$ is a geodesic, shear free,
expanding as well as rotating null vector of both
$g_{ab}^{\rm KN}$ as well as $g_{ab}^{\rm KNV}$ and given in (2.5)
and $g_{ab}^{\rm KN}$ is the Kerr-Newman metric
(3.4) with $m=e$ = constant. This null vector $\ell_a$ is
one of the double repeated principal
null vectors of the Weyl tensor of $g_{ab}^{\rm KN}$.

It appears that the rotating Kerr-Newman geometry may be regarded
as joining smoothly with the rotating Vaidya geometry at its null
radiative boundary, as shown by Glass and Krisch [7] in the case
of Schwarzschild geometry joining to the non-rotating Vaidya
space-time. The Kerr-Schild form (3.19) will recover that of
Xanthopoulos [8] $g'_{ab}=g_{ab}+\ell_a\ell_b$, when
$Q(u,r,\theta) \rightarrow 1/2$ and that of Glass and Krisch [7]
$g'_{ab}=g_{ab}^{\rm Sch}-\{2f(u)/r\}\ell_a\ell_b$ when $e=a=0$
for non-rotating Schwarzschild background space. Thus, one can
consider the Kerr-Schild form (3.19) as the extension of those of
Xanthopoulos as well as Glass and Krisch. When we set $a=0$, this
rotating Kerr-Newman-Vaidya solution (3.15) will recover to
non-rotating Reissner-Nordstrom-Vaidya solution with the
Kerr-Schild form $g_{ab}^{\rm RNV}=g_{ab}^{\rm RN} -
\{2f(u)/r\}\ell_a\ell_b$, which is still a generalization of
Xanthopoulos as well as Glass and Krisch in the charged
Reissner-Nordstrom solution. It is worth mentioning that the new
solution (3.15) cannot be considered as a bimetric theory as
$g_{ab}^{\rm KNV}\neq{1\over2}(g_{ab}^{\rm KN}+g_{ab}^{\rm V})$.

To interpret the Kerr-Newman-Vaidya solution as a black hole
during the early inflationary phase of rotating Vaidya null
radiating universe {\it i.e.}, the Kerr-Newman black hole embedded
into rotating Vaidya null radiating background space, we can write
the Kerr-Schild form (3.19) as
\begin{equation}
g_{ab}^{\rm KNV}=g_{ab}^{\rm V} +2Q(r,\theta)\ell_a\ell_b
\end{equation}
where
\begin{equation}
Q(r,\theta) =-(rm - e^2/2)R^{-2}.
\end{equation}
Here, the constants $m$ and $e$ are the mass and the charge of
Kerr-Newman black hole, $g_{ab}^{\rm V}$ is the rotating
Vaidya null radiating black hole obtained above when $e(u)$ sets
to zero in (3.4) and $\ell_a$ is the geodesic null vector
given in (2.5) for both $g_{ab}^{\rm KNV}$ and $g_{ab}^{\rm V}$.
When we set $f(u)=a=0$, $g_{ab}^{\rm V}$ will recover the flat
metric, then  $g_{ab}^{\rm KNV}$ becomes the original Kerr-Schild
form written in spherical symmetric flat background.

These two Kerr-Schild forms (3.19) and (3.21) certainly confirm
that the metric  $g_{ab}^{\rm KNV}$  is a solution of Einstein's
field equations since the background rotating metrics $g_{ab}^{\rm
KN}$ and  $g_{ab}^{\rm V}$ are solutions of Einstein's equations.
They both possess different stress-energy tensors $T_{ab}^{(\rm
e)}$ and $T_{ab}^{(\rm n)}$ given in (3.18) and (3.17)
respectively. Looking at the Kerr-Schild form (3.21), the
Kerr-Newman-Vaidya black hole can be treated as a generalization
of Kerr-Newman black hole by incorporating Visser's suggestion
[26] that {\it Kerr-Newman black hole embedded in an axisymmetric
cloud of matter would be of interest}.

\vspace{.15in}

{\it (iv) Rotating de Sitter solution}
\vspace*{.15in}

Here we shall first convert the standard de Sitter cosmological
universe into a rotating de Sitter space. So that the rotating
Kerr-Newman-Vaidya solution can be embedded into the rotating de
Sitter cosmological universe as Kerr-Newman-Vaidya-de Sitter
solution. It means that one must a rotating solution in oder to 
embed into a rotating one, leading to a feasible solution . For 
this purpose, one can choose the Wang-Wu function $q_n(u)$ given 
in (3.9) as
\begin{eqnarray}
\begin{array}{cc}
q_n(u)=&\left\{\begin{array}{ll}
\Lambda^*/6, &{\rm when}\;\;n=3\\
0, &{\rm when }\;\;n\neq 3
\end{array}\right.
\end{array}
\end{eqnarray}
such that the mass function becomes
\begin{equation}
M(u,r)={\Lambda^*\,r^3\over 6}.
\end{equation}
Then the line element of the rotating de Sitter metric will take  the
following form
\begin{eqnarray}
ds^2&=&\Big\{1-{\Lambda^*\,r^4\over3\,R^2}\Big\}\,du^2 +2du\,dr
-R^2d\theta^2 \cr &&+\frac{2a\Lambda^*\,r^4}{3\,R^2}\,{\rm
sin}^2\theta\,du\,d\phi -2a\,{\rm sin}^2\theta\,dr\,d\phi \cr
&&-\Big\{(r^2+a^2)^2 -\Delta^*a^2\,{\rm
sin}^2\theta\Big\}\,R^{-2}{\rm sin}^2\theta\,d\phi^2,
\nonumber \\
\end{eqnarray}
where $R^2=r^2+a^2{\rm cos}^2\theta$ and
$\Delta^*=r^2-{\Lambda^*\,r^4}/3+a^2$. This corresponds to the
rotating de Sitter solution for $\Lambda^*>0$, and to the anti-de
Sitter one for $\Lambda^*<0$. In general $\Lambda^*$ denotes the
cosmological constant of the de Sitter space. The rotating
cosmological de Sitter space possesses an energy momentum tensor
\begin{eqnarray}
T_{ab} =2\,\rho^*\,\ell_{(a}\,n_{b)}
+2\,p\,m_{(a}\overline m_{b)},
\end{eqnarray}
where
\begin{eqnarray*}
\rho^*={\Lambda^*r^4\over {K\,R^2\,R^2}}, \:\:\:\: p=
{-\Lambda^*r^2\over{K\,R^2\,R^2}}(r^2+2a^2\,{\rm cos}^2\theta).
\end{eqnarray*}
are related to the density and the pressure  of the cosmological
matter field which is, however not a perfect fluid. Then the
changed NP quantities are
\begin{eqnarray}
&&\gamma=-{1\over{2\overline R\,R^2}}\,\Big\{\Big(1-{2\over
3}\Lambda^*r^2\Big)r\,\overline R+\Delta^*\Big\}, \cr &&
\phi_{11}=
-{1\over {2\,R^2\,R^2}}\Lambda^*r^2a^2\,{\rm cos}^2\theta, \\
&&\psi_2={1\over{3\overline R\,\overline
R\,R^2}}\Lambda^*r^2a^2{\rm
cos}^2\theta, \\
&&\Lambda={\Lambda^*r^2\over{6R^2}}.
\end{eqnarray}
This means that in rotating de Sitter cosmological universe, the
$\Lambda^*$ is coupling with the rotational parameter $a$. From
these NP quantities we clearly observe that the rotating de Sitter
cosmological metric is a Petrov type $D$ gravitational field
$\psi_2 \neq 0$, whose one of the repeated principal null vectors,
$\ell_a$ is geodesic, shear free, expanding as well as non-zero
twist. The metric (3.25) has singularities at the values of $r$
for which $\Delta^*=0$ having four roots $r_{++}$, $r_{+-}$,
$r_{-+}$ and $r_{--}$. The singularity at
$r_{++}=+\surd[(1/2\Lambda^*)\{3+\surd(9+12a^2\Lambda^*)\}]$ might
represent the apparent singularity for the rotating de Sitter
space (3.25). When $a=0$ for the non-rotating de Sitter space,
this will reproduce the result 
$r_{++}=3^{1/2}\Lambda^{*-1/2}$, discussed by Gibbon and Hawking
[28]. The surface gravity of the singularity at $r=r_{++}$ can be
written as
\begin{eqnarray*}
{\cal K}=-\frac{r_{++}}{12R^2}\Big\{9-\sqrt{(9+12a^2\Lambda^*)}\,\Big\}.
\end{eqnarray*}
The angular velocity for the singularity is
\begin{eqnarray*}
\Omega_{\rm H}=\frac{a\Lambda^{*}r^4}{3(r^2+a^2)^2}\Big|_{r=r_{++}}.
\end{eqnarray*}
If we set the rotational parameter $a=0$, we will recover the
non-rotating de Sitter metric [28], which is a solution of the
Einstein's equations for an empty space with $\Lambda \equiv
(1/24)g_{ab}R^{ab}=\Lambda^*/6$ or constant curvature and
$\rho^*=-p$. However, it is observed that the rotating de Sitter
universe (3.25) is {\sl non-empty} and {\sl non-constant
curvature}. It certainly describes a stationary rotating spherical
symmetric solution representing Petrov type $D$ spacetime with the Weyl
scalar $\psi_2$ (3.28). So it
is noted that to the best of the present author's knowledge, this
rotating de Sitter metric has not been seen derived  before.

\vspace*{.15in}

{\sl (v) Kerr-Newman-Vaidya-de Sitter solution}

\vspace*{.15in}

Now we shall embed the Kerr-Newman-Vaidya solution (3.15) into the
rotating de Sitter solution (3.25) by choosing the Wang-Wu
function as
\begin{eqnarray}
\begin{array}{cc}
q_n(u)=&\left\{\begin{array}{ll}
m+f(u),&{\rm when }\;\;n=0\\
-e^2/2, &{\rm when }\;\;n=-1\\
\Lambda^*/6, &{\rm when}\;\;n=3\\
0, &{\rm when }\;\;n\neq 0, -1, 3,
\end{array}\right.
\end{array}
\end{eqnarray}
where $m$ and $e$ are constants and $f(u)$ is related with the
mass of rotating Vaidya solution (3.15). Thus, we have the
mass function
\begin{eqnarray*}
M(u,r)=m+f(u)-{e^2\over 2r}+{\Lambda^*\,r^3\over 6}
\end{eqnarray*}
and other quantities are obtained from (3.6) and (2.13) as
\begin{eqnarray}
&&\rho^* = {1\over K\,R^2\,R^2}\Big(e^2 + \Lambda^*r^4\Big),
\\&&p= {1\over K\,R^2\,R^2}\Big\{e^2 - \Lambda^*r^2(r^2 +2a^2{\rm
cos}\theta)\Big\}, \cr &&\mu^* ={-r\over
K\,R^2\,R^2}\Big\{2\,r\,f(u)_{,u}+a^2{\rm
sin}^2\theta\,f(u)_{,uu}\Big\}, \cr &&\omega ={-i\,a\,{\rm
sin}\,\theta\,\over{\surd 2\,K\,\overline R\,R^2}}\,f(u)_{,u}, \cr
&&\Lambda\equiv \frac{1}{24}g^{ab}R_{ab}={\Lambda^*r^2 \over
6\,R^2}, \\ &&\gamma={1\over{2\overline
R\,R^2}}\,\Big[\Big\{r-m-f(u)
-{2\Lambda^*r^3\over3}\Big\}\overline R-\Delta^*\Big],\nonumber
\end{eqnarray}
and $\phi_{11}$, $\phi_{12}$, $\phi_{22}$ can be obtained from
equations (3.32) with (2.12). The Weyl scalars take the following
form
\begin{eqnarray}
&\psi_2=&{1\over\overline R\,\overline R\,R^2}\Big[e^2
-R\{m+f(u)\}\cr &&+{\Lambda^*r^2\over 3}a^2{\rm
cos}^2\theta\Big],\cr &\psi_3 =&{-i\,a\,{\rm sin}\theta\over
2\surd 2\overline
R\,\overline R\,R^2} \Big\{(4\,r+\overline R)f(u)_{,u}\Big\}, \\
&\psi_4 =&{{a^2r\,{\rm sin}^2\theta}\over 2\overline R\,\overline
R\,R^2\,R^2}\,\Big\{R^2f(u)_{,uu}-2rf(u)_{,u}\Big\}. \nonumber
\end{eqnarray}
That is, the rotating Kerr-Newman-Vaidya-de Sitter solution
will take the line element as follows
\begin{eqnarray}
ds^2&=&\Big[1-R^{-2}\Big\{2r\Big(m+f(u)\Big)+{\Lambda^*r^4\over
3}-e^2\Big\}\Big]du^2 \cr &&+2du\,dr +2aR^{-2}\Big\{2r\Big(m
+f(u)\Big)+{\Lambda^*\,r^4\over 3} \cr &&-e^2\Big\}\,{\rm sin}^2
\theta\,du\,d\phi -2a\,{\rm sin}^2\theta\,dr\,d\phi -R^2d\theta^2
\cr && -\Big\{(r^2+a^2)^2 -\Delta^*a^2\,{\rm
sin}^2\theta\Big\}\,R^{-2}{\rm
sin}^2\theta\,d\phi^2, \nonumber \\
\end{eqnarray}
where $\Delta^*=r^2-2r\{m+f(u)\}-\Lambda^*\,r^4/3+a^2+e^2$. Here
$m$ and $e$ are the mass and the charge of Kerr-Newman solution,
$a$ is the  non-zero rotation parameter and $f(u)$ represents the
mass function of rotating Vaidya null radiating fluid. The metric 
(3.34) will describe a cosmological black holes with the horizons
at the values of $r$ for which $\Delta^*=0$ having four roots 
$r_{++}$, $r_{+-}$, $r_{-+}$ and $r_{--}$ given in appendix (A5) 
and (A6). The first three values will describe respectively the 
event horizon, the Cauchy horizon and the cosmological horizon. 
The surface gravity of the horizon at
$r=r_{++}$ is
\begin{eqnarray*}
{\cal K}=-\Big[\,\frac{1}{rR^2}\Big\{r\Big(r-m-f(u)
-\frac{\Lambda^*r^3}{6}\Big)+\frac{e^2}{2}\Big\}\Big]_{r=r_{++}}
\end{eqnarray*}
and the entropy of the horizon is obtained as
\begin{eqnarray*}
{\cal S}=\pi\Big\{r^{2}+a^2\Big\}_{r=r_{++}}.
\end{eqnarray*}
The angular velocity of the horizon is found as 
\begin{eqnarray*}
\Omega_{\rm H}=\frac{a[2r\{m+f(u)\}+(\Lambda^{*}r^4/3)-
e^2(u)]}{(r^2+a^2)^2}\Big|_{r=r_{++}}.
\end{eqnarray*}

In this
rotating solution (3.34), the Vaidya null fluid is interacting
with the non-null electromagnetic field on the de Sitter
cosmological space. Thus, the total energy momentum tensor (EMT)
for the rotating solution (3.34) takes the following form:
\begin{eqnarray}
T_{ab} = T^{(n)}_{ab} +T^{(\rm E)}_{ab}+T^{(\rm C)}_{ab},
\end{eqnarray}
where the EMTs for the rotating null fluid, the
electromagnetic field and cosmological matter field are given
respectively
\begin{eqnarray}
&&T^{(\rm n)}_{ab}= \mu^*\,\ell_a\,\ell_b +
2\,\omega\,\ell_{(a}\,\overline
m_{b)}+2\,\overline\omega\,\ell_{(a}\,m_{b)}, \cr &&T^{(\rm
E)}_{ab} =4\,\rho^{*(\rm E)}\{\ell_{(a}\,n_{b)} +m_{(a}\overline
m_{b)}\}, \\ &&T^{(\rm C)}_{ab}=2\{\rho^{*(\rm
C)}\,\ell_{(a}\,n_{b)} +2\,p^{(\rm C)}\,m_{(a}\overline
m_{b)}\},\nonumber
\end{eqnarray}
where $\mu^*$ and $\omega$ are given in (3.32) and
\begin{eqnarray}
&&\rho^{*(\rm E)}=p^{(\rm E)}= {e^2\over {K\,R^2\,R^2}}, \:\:\:
\rho^{*(\rm C)}= {\Lambda^*r^4\over {K\,R^2\,R^2}}, \cr &&p^{(\rm
C)}= -{\Lambda^*r^2\over{K\,R^2\,R^2}}\Big(r^2+2a^2\,{\rm
cos}^2\theta\Big).
\end{eqnarray}
Now, for future use we shall, without loss of generality, have a
decomposition of the Ricci scalar $\Lambda$, given in (3.32), as
\begin{equation}
\Lambda = \Lambda^{(\rm E)} + \Lambda^{(\rm C)}
\end{equation}
where $\Lambda^{(\rm E)}$ is the {\sl zero} Ricci scalar for the
electromagnetic  field and  $\Lambda^{(\rm C)}$ is the {\sl
non-zero} cosmological Ricci scalar with $\Lambda^{(\rm C)}=
(\Lambda^*r^2/6R^2)$. The appearance of $\omega$ shows that the
Vaidya null fluid is rotating as the expression of $\omega$ in
(3.36) involves the rotating parameter $a$ coupling with $\partial
f(u)/\partial u$ -- both are non-zero quantities for a rotating
Vaidya null radiating universe.

For non-rotating spacetime with $a=0$, the total energy momentum
tensor (3.35), will become the  following form
\begin{equation}
T_{ab} =T^{(\rm n)}_{ab}+T^{(\rm E)}_{ab} + \Lambda^*g_{ab}
\end{equation}
where $T^{(\rm E)}_{ab}$ is the energy momentum tensor for
non-null electromagnetic field existing in the non-rotating
Reissner-Nordstrom-de Sitter metric $g_{ab}$ and $T^{(\rm
n)}_{ab}=\mu^*\,\ell_a\,\ell_b$ is the energy momentum tensor for
the standard non-rotating Vaidya null fluid. The energy momentum
tensor (3.39) may be considered as Guth's modification of $T_{ab}$
[29] in non-stationary charged Vaidya-de Sitter universe.

This rotating Kerr-Newman-Vaidya-de Sitter metric (3.34) can be
written in a Kerr-Schild form on the de Sitter background as
\begin{equation}
g_{ab}^{\rm KNVdS}=g_{ab}^{\rm VdS} +2Q(u,r,\theta)\ell_a\ell_b
\end{equation}
where $Q(u,r,\theta) =-\{r\,m-e^2/2\}R^{-2}$, and the vector
$\ell_a$ is a geodesic, shear free, expanding as well as non-zero
twist null vector of both $g_{ab}^{\rm VdS}$ as well as
$g_{ab}^{\rm KNVdS}$ and given in (2.5). We can also write this
solution (3.34) in another Kerr-Schild form on the Kerr-Newman
background as
\begin{equation}
g_{ab}^{\rm KNVdS}=g_{ab}^{\rm KN} +2Q(u,r,\theta)\ell_a\ell_b
\end{equation}
where $Q(u,r,\theta) =-\{rf(u)+\Lambda^*r^4/6\}R^{-2}$. These two
Kerr-Schild forms (3.40) and (3.41) certainly assure that the
metric $g_{ab}^{\rm KNVdS}$  is a solution of Einstein's field
equations, since the background rotating metrics  $g_{ab}^{\rm
KN}$ and $g_{ab}^{\rm VdS}$ are both solutions of Einstein's field
equations. They have different stress-energy tensors $T_{ab}^{(\rm
E)}$, $T_{ab}^{(\rm n)}$ and $T_{ab}^{(\rm C)}$ given in (3.36).

From the rotating solution (3.34), one can recover (i) the
Kerr-Newman-de Sitter when $f(u)=0$, (ii) the rotating charged
Vaidya-de Sitter null radiating black hole if $m=0$, (iii) the
rotating Kerr-Newman-Vaidya metric (3.15) when $\Lambda^*=0$. If
one sets $f(u)=m=e=0$, one will get the rotating de Sitter
solution (3.25). All these rotating solutions mentioned here will
be of interest to study the physical properties of embedded
rotating solutions. One can also find that the rotating
Kerr-Newman-de Sitter solution (3.34) with $f(u)=0$ and its
non-stationary extension ($f(u) \neq 0$) are different from the
ones derived by Carter [30], Mallett [31] and Xu [32] in the terms
involving $\Lambda^*$.

\vspace*{.15in}

{\sl (vi) Kerr-Newman-monopole solution}

\vspace*{.15in}

Now we shall give another example of embedded solution of the
Kerr-Newman black hole into the rotating monopole universe by
choosing the Wang-Wu function as
\begin{eqnarray}
\begin{array}{cc}
q_n(u)=&\left\{\begin{array}{ll}
m,&{\rm when }\;\;n=0\\
b/2, &{\rm when }\;\;n=1\\
-e^2/2, &{\rm when }\;\;n=-1\\
0, &{\rm when }\;\;n\neq 0, \pm 1,
\end{array}\right.
\end{array}
\end{eqnarray}
where $m$ and $e$ are constants and $b$ is identified as monopole
constant [9]. Thus, the mass function becomes
\begin{eqnarray*}
M(u,r)=m+{r\,b\over 2}-{e^2\over 2r}
\end{eqnarray*}
and other quantities are obtained from (3.6) and (2.13) as
\begin{eqnarray}
&&\mu^* =\omega=0, \nonumber \\\ &&\rho^* = {1\over
K\,R^2\,R^2}\Big(e^2 + b\,r^2\Big), \nonumber \\\ &&p= {1\over
K\,R^2\,R^2}\Big\{e^2 - b\,a^2{\rm cos}\theta)\Big\}, \\\
&&\Lambda={b\over 12\,R^2}, \\
&&\psi_2={1\over\overline R\,\overline RR^2}\Big[e^2 -Rm-+{b\over
6}\Big(\overline R\,\overline R-2ira{\rm cos}\theta\Big)\Big]. \cr
&&\gamma={1\over{2\overline R\,R^2}}\,\left[\{r(1-b)-m\}\overline
R-\Delta^*\right]. \nonumber
\end{eqnarray}
Then the embedded solution will be as follows
\begin{eqnarray}
ds^2&=&[1-R^{-2}(2rm+b\,r^2-e^2)]\,du^2+2du\,dr \cr
&&+2aR^{-2}(2rm+b\,r^2-e^2)\,{\rm sin}^2 \theta\,du\,d\phi \cr
&&-2a\,{\rm sin}^2\theta\,dr\,d\phi -R^2d\theta^2 -\{(r^2+a^2)^2
\cr &&-\Delta^*a^2\,{\rm sin}^2\theta\}\,R^{-2}{\rm
sin}^2\theta\,d\phi^2,
\end{eqnarray}
where $\Delta^*=r^2(1-b)-2rm+a^2+e^2$. Here $m$ and $e$ are the
mass and the charge of Kerr-Newman solution, $a$ is the non-zero
rotation parameter and $b$ represents the monopole constant. In
this rotating solution (3.44), the matter field describing
monopole particles is interacting with the non-null
electromagnetic field. Thus, the total energy momentum tensor
(EMT) for the rotating solution (3.45) takes the following form:
\begin{eqnarray}
T_{ab} = T^{(\rm E)}_{ab}+T^{(\rm m)}_{ab},
\end{eqnarray}
where the EMTs for the monopole matter field and electromagnetic
field are given respectively
\begin{eqnarray}
&&T^{(\rm m)}_{ab}=2\{\rho^{*(\rm m)}\,\ell_{(a}\,n_{b)}
+2\,p^{(\rm m)}\,m_{(a}\overline m_{b)}\}, \cr &&T^{(\rm E)}_{ab}
=4\,\rho^{*(\rm E)}\{\ell_{(a}\,n_{b)} +m_{(a}\overline m_{b)}\},
\end{eqnarray} where
\begin{eqnarray}
&&\rho^{*(\rm E)}=p^{(\rm E)}= {e^2\over {K\,R^2\,R^2}}, \:\:\:
\rho^{*(\rm m)}= {b\,r^2\over {K\,R^2\,R^2}}, \cr &&p^{(\rm m)}=
-{1\over{K\,R^2\,R^2}}\,b\,a^2\,{\rm cos}^2\theta.
\end{eqnarray}
Accordingly, for future use one can have a decomposition of
$\Lambda$ given in (3.43) as
\begin{equation}
\Lambda = \Lambda^{(\rm E)} + \Lambda^{(\rm m)}
\end{equation}
where $\Lambda^{(\rm E)}$ is the {\sl zero} Ricci scalar for the
electromagnetic  field and  $\Lambda^{(\rm m)}$ is the {\sl
non-zero} monopole Ricci scalar with $\Lambda^{(\rm m)}=
b/(12R^2)$. One has also seen the interaction of the rotating
parameter $a$ with the monopole constant $b$ in the expression of
$p^{(\rm m)}$, which makes different between the rotating as well
as non-rotating monopole solutions.

The roots of the equation $\Delta^*=0$ are found as
\begin{equation}
r_{\pm}=\frac{1}{(1-b)}\Big[m\pm \sqrt{
\{m^2-(1-b)(a^2+e^2)\}}\,\Big].
\end{equation}
From this we observe that the value of $b$ must lie in $0<b<1$, 
with the horizons at $r=r_{\pm}$. The surface gravity of the
horizon at $r=r_+$ is
\begin{eqnarray*}
{\cal K}=-\frac{1}{r_+R^2}\Big[r_+\Big\{r_+
\Big(1-\frac12\,b\Big)-m\Big\}+\frac{e^2}{2}\Big].
\end{eqnarray*}
The entropy and the angular velocity of the horizon are found as
\begin{eqnarray*}
&&{\cal S}=\pi\{r^{2}_{+}+a^2\}, \cr
&&\Omega_{\rm H}=\frac{a\{2rm+br-e^2\}}{(r^2+a^2)^2}\Big|_{r=r_{+}}.
\end{eqnarray*}

We have also found that the solution (3.45) represents Petrov type
$D$ with the Weyl scalar $\psi_2$ given in (3.44) whose repeated
principal null vector $\ell_a$ is shear free, rotating and
non-zero twist. This Kerr-Newman-monopole solution (3.45) can be
written in a Kerr-Schild form on the rotating monopole background
as
\begin{equation}
g_{ab}^{\rm KNm}=g_{ab}^{\rm m}+2Q(r,\theta)\ell_a\ell_b
\end{equation}
where $Q(r,\theta) =-(rm-e^2/2)R^{-2}$, and the vector $\ell_a$ is
a geodesic, shear free, expanding as well as non-zero twist null
vector of both $g_{ab}^{\rm m}$ as well as $g_{ab}^{\rm KNm}$ and
given in (2.5). We can also express this solution (3.45) in
another Kerr-Schild ansatz on the Kerr-Newman background as
\begin{equation}
g_{ab}^{\rm KNm}=g_{ab}^{\rm KN}+2Q(r,\theta)\ell_a\ell_b
\end{equation}
where $Q(r,\theta) =-(b\,r/4)R^{-2}$. These two Kerr-Schild forms
(3.51) and (3.52) assure that the metric $g_{ab}^{\rm KNm}$ is a
solution of Einstein's field equations, since the background
rotating metrics  $g_{ab}^{\rm KN}$ and $g_{ab}^{\rm m}$ are both
solutions of Einstein's field equations. They have different
stress-energy tensors $T_{ab}^{(\rm E)}$ and $T_{ab}^{(\rm m)}$
given in (3.47).

From the rotating solution (3.45), we can recover (i) the
Kerr-Newman metric when $b=0$, (ii) the rotating charged monopole
solution if $m=0$, (iii) if one sets $m=e=0$, one will get the
rotating monopole solution. All these rotating solutions mentioned
here will be of interest to study the physical properties of
embedded rotating solutions. (iv) For $a=0$, the metric (3.45)
will reduce to the non-rotating Reissner-Nordstrom-monopole
solution. (v) If $m=e=a=0$, it will recover the non-rotating
monopole solution [9].

\begin{center}
{\bf 4. HAWKING'S RADIATION ON THE VARIABLE CHARGED
BLACK HOLES}
\end{center}
\setcounter{equation}{0}
\renewcommand{\theequation}{4.\arabic{equation}}

In this section, as a part of discussion of the physical
properties of the embedded solutions (3.15), (3.34) and (3.45), we
shall discuss describe a scenario which is capable of avoiding the
formation of negative mass naked singularity during Hawking
radiation process in spacetime metrics, describing the life style
of a rotating embedded radiating black hole. The formation of
negative mass naked singularities in classical spacetime metrics
is being shown in [1] after the complete evaporation of the masses
of {\sl non-embedded} rotating Kerr-Newman and non-rotating
Reissner-Nordstrom, black holes due to Hawking electrical
radiation.

Here we shall clarify two similar nomenclatures having different
meaning like Hawking's radiation and Vaidya null radiation.
Hawking's radiation is continuous radiation of energy from the
black hole body thereby leading to the change in mass [1-5],
whereas the Vaidya null radiation means that the stress-energy
momentum tensor describing the gravitation in Vaidya spacetime
metric is a null radiating fluid [33]. So Vaidya null radiation
does not have any direct relation with Hawking's radiation of
black holes [10].

\vspace{0.15in}

{\it (i) Variable-charged Kerr-Newman-Vaidya black hole}

\vspace{0.15in}

As mentioned earlier in the introduction,
by electrical radiation of a charged black hole we mean the
variation of the charge $e$ with respect to the coordinate $r$
in the stress-energy momentum tensor of electromagnetic field.
This variation of $e$ will certainly lead different forms or functions of
the stress-energy tensor from that of Kerr-Newman-Vaidya black
hole. To observe the change in the mass of black hole in
the spacetime metric, one has to consider a different form or
function of stress-energy tensor of a particular black hole. That
is, in order to incorporate the Hawking's radiation in
this black hole (3.15), we must have a different stress-energy
tensor as the Kerr-Newman-Vaidya black hole with $T_{ab}$
(3.16) does not have any direct Hawking's radiation effects.
The consideration of different forms of stress-energy-momentum
tensor in the study of Hawking's radiation effect in classical
spacetime metrics here is followed from Boulware's suggestion
[5] mentioned in introduction above.

It is noted that the Kerr-Newman-Vaidya black hole, describing the
Kerr-Newman black hole embedded into the rotating Vaidya null
radiating universe (3.4), is quite different from the standard
Kerr-Newman black hole. That is, (i) the Kerr-Newman-Vaidya black
hole is algebraically special in Petrov classification with the
Weyl scalars $\psi_{0}=\psi_{1}=0$, $\psi_{2}\neq \psi_{3}\neq
\psi_{4}\neq 0$, where as the standard Kerr-Newman black hole is
Petrov type $D$ with $\psi_{0}=\psi_{1}=\psi_{3}=\psi_{4}=0$,
$\psi_{2} \neq 0$. (ii) the Kerr-Newman-Vaidya black hole
possesses the total energy-momentum tensor (3.16), representing
interaction of the null radiating fluid $T^{(\rm n)}_{ab}$ with
the electromagnetic field $T^{(\rm E)}_{ab}$, i.e., the charged
null radiating fluid; however the energy-momentum tensor of the
Kerr-Newman black hole is that of electromagnetic field $T^{(\rm
E)}_{ab}$, simply a charged black hole. (iii) the
Kerr-Newman-Vaidya black hole is a non-stationary extension of the
stationary Kerr-Newman black hole. Due to the above differences
between the two black holes, it is worth studying the physical
properties of the embedded black hole. Hence, it is hoped that the
study of Hawking electrical radiation in the rotating
Kerr-Newman-Vaidya black hole will certainly lead to a different
physical feature than that of {\sl non-embedded} Kerr-Newman black
hole.

Thus, we consider the line element of a variable-charged
Kerr-Newman-Vaidya black hole with respect to the coordinate $r$
as follows (by changing $m$ to $M$ for better understanding):
\begin{eqnarray}
d s^2&=&[1-R^{-2}\{2r(M+f(u))- e^2(r)\}]\,du^2 \cr &&+2du\,dr
+2aR^{-2}\{2r(M+f(u)) \cr &&-e^2(r)\}\,{\rm sin}^2
\theta\,du\,d\phi -2a\,{\rm sin}^2\theta\,dr\,d\phi \cr
&&-R^2d\theta^2 -\{(r^2+a^2)^2 \cr &&-\Delta^*a^2\,{\rm
sin}^2\theta\}\,R^{-2}{\rm sin}^2\theta\,d\phi^2,
\end{eqnarray}
where $\Delta^*=r^2-2r\{M+f(u)\}+a^2+e^2(r)$. Then, with the
variable charge $e(r)$ in the above metric, there will be a
different energy-momentum tensor from that of the original
Kerr-Newman-Vaidya metric (3.15). Accordingly, we have to
calculate the Ricci and Weyl scalars from the field equations with
the functions $e(r)$. However, instead of calculating directly
from the field equations, one can use the solutions given in (3.6)
and (3.8) to obtain these scalars. For the completeness we present
them as follows:
\begin{eqnarray}
&\phi_{11}=&{1\over
{2\,R^2\,R^2}}\,\Big\{e^2(r)-2r\,e(r)e(r)_{,r}\Big\}
\cr  &&+{1\over 4R^2}\,\Big\{e^2(r)_{,r}+e(r)\,e(r)_{,rr}\Big\}, \\
&\phi_{22}=&-{r\over 2\,R^2\,R^2}\Big\{2\,r\,f(u)_{,u}+a^2{\rm
sin}^2\theta\,f(u)_{,uu}\Big\}, \cr &\phi_{12}=&{i\,a\,{\rm
sin}\,\theta\,\over{\surd 2\,K\,\overline R\,R^2}}\,f(u)_{,u},
\nonumber \\ &\Lambda =& -{1\over
12\,R^2}\,\Big\{e^2(r)_{,r}+e(r)\,e(r)_{,rr}\Big\}, \\
&\psi_2=&{1\over\overline R\,\overline
R\,R^2}\Big[e^2(r)-R\{M+f(u)\} \cr &&+{1\over 6}\,\overline
R\,\overline R\,\Big\{e^2(r)_{,r}
+e(r)\,e(r)_{,rr}\Big\} \cr &&-\overline R\,e(r)\,e(r)_{,r}\Big], \\
&\psi_3=&{-i\,a\,{\rm sin}\theta\over 2\surd 2\overline
R\,\overline R\,R^2}\Big\{(4\,r+\overline R)f(u)_{,u}\Big\}, \cr
&\psi_4=&{{a^2r\,{\rm sin}^2\theta}\over 2\overline R\,\overline
R\,R^2\,R^2}\,\Big\{R^2f(u)_{,uu}-2rf(u)_{,u}\Big\}. \nonumber
\end{eqnarray}
From these we observe that the variable charge $e(r)$ has the
effect only on the scalars $\phi_{11}$, $\Lambda$ and $\psi_2$.
However, the non-vanishing of $\phi_{22}$, $ \phi_{12}$,
$\psi_3$,and $\psi_4$ indicate the different physical feature of
non-statio\\nary Kerr-Newman-Vaidya black hole from the stationary
standard Kerr-Newman black hole. One important feature of the
field equations corresponding to the metric (4.1) with $e(r)$ is
that the expressions for $\phi_{11}$ and $\Lambda$ do not involve
the Vaidya mass function $f(u)$. So it suggests the possibility to
study the Hawking electrical radiation in this non-stationary
black hole.

Now for an electromagnetic field, the Ricci scalar $\Lambda\equiv
(1/24)g^{ab}R_{ab}$ (4.3) has to vanish leading to the solution
\begin{equation}
e^2(r)=2rm_1 + C
\end{equation}
where $m_1$ and $C$ are real constants. Substituting this
$e^2(r)$ in (4.2) and (4.4), and after identifying the
constant $C\equiv e^2$, we obtain
\begin{eqnarray}
&&\phi_{11}={e^2\over {2\,R^2\,R^2}}, \\
&&\psi_2={1\over\overline R\,\overline
R\,R^2}\Big[e^2-R\{M-m_1+f(u)\}\Big]. \nonumber \\
\end{eqnarray}
and $\phi_{12}$, $\phi_{22}$, $\psi_3$, and $\psi_4$ are remained
unaffected as in (4.2) and (4.4) above. Accordingly, the Maxwell
scalar $\phi_1$ with constant $e$ becomes
\begin{eqnarray}
\phi_{1}={e\over {\surd 2\,\overline R\,\overline R}},
\end{eqnarray}
which is the same Maxwell scalar of the Kerr-\\Newman-Vaidya
metric (3.15) if once written out from (3.12). From the Weyl scalar
(4.7) we have clearly seen {\sl a change in the mass $M$} by some
constant quantity $m_1$ (say) for the first  electrical
radiation in the embedded black hole. Then the total mass of
Kerr-Newman-Vaidya black hole becomes $(M-m_1)+f(u)$ and the
metric takes the form
\begin{eqnarray}
ds^2&=&[1-R^{-2}\{2r(M-m_1 +f(u))-e^2\}]\,du^2 \cr &&+2du\,dr
+2aR^{-2}\{2r(M-m_1+f(u)) \cr &&-e^2\}\,{\rm
sin}^2\theta\,du\,d\phi -2a\,{\rm sin}^2\theta\,dr\,d\phi
-R^2d\theta^2 \cr &&-\{(r^2+a^2)^2 -\Delta^*a^2\,{\rm
sin}^2\theta\}\,R^{-2}{\rm
sin}^2\theta\,d\phi^2, \nonumber \\
\end{eqnarray}
where $\Delta^*=r^2-2r\{M-m_1+f(u)\}+a^2+e^2$. Since the Maxwell
scalar (4.8) remains the same for the first Hawking radiation, we
again consider the charge $e$ to be a function of $r$ for the
second radiation in the metric (4.9) with the mass $M-m_1+f(u)$.
This will certainly lead, by the Einstein-Maxwell field equations
with the vanishing of $\Lambda$ to reduce another quantity $m_2$
(say) from the total mass, {\it i.e.} the mass becomes
$M-(m_1+m_2)+f(u)$ after the second electrical radiation. Here
again, we observe that the Maxwell scalar $\phi_1$ remains the
same form and also there is no effect on the Vaidya mass function
$f(u)$ after the second radiation. Thus, if we consider $n$
radiations, every time taking the charge $e$ to be a function of
$r$, the Einstein's field equations will imply that the total mass
of the black hole will take the form
\begin{equation}
{\cal M}=M-(m_1 + m_2 + m_3 + m_4 + . . .+ m_n)+f(u)
\end{equation}
without affecting the mass function $f(u)$. Taking Hawking's
radiation of charged black hole embedded in the rotating Vaidya
null radiating space, one might expect that the mass $M$ may be
radiated away, just leaving $M-(m_1 + m_2 + m_3 + m_4 + . . .+
m_n)$ equivalent to the Planck mass of about $10^{-5}\,{\rm g}$
and $f(u)$ untouched; that is, $M$ may not be exactly equal to
$(m_1 + m_2 + m_3 + m_4 +  . . . + m_n)$, but has a difference of
about a Planck-size mass. Otherwise, the mass $M$ may be
evaporated completely after continuous radiation, when $M=(m_1 +
m_2 + m_3 +m_4+ . . .+ m_n)$, just leaving the Vaidya mass
function $f(u)$ and the electric charge $e$ only. Thus, we can
show this situation of the black hole in the form of a classical
spacetime metric as
\begin{eqnarray}
d s^2&=&[1-R^{-2}\{2rf(u)-e^2\}]\,du^2 +2du\,dr\cr
&&+2aR^{-2}\{2rf(u)-e^2\}\,{\rm sin}^2 \theta\,du\,d\phi \cr
&&-2a\,{\rm sin}^2\theta\,dr\,d\phi -R^2d\theta^2 -\{(r^2+a^2)^2
\cr &&-\Delta^*a^2\,{\rm sin}^2\theta\}\,R^{-2}{\rm
sin}^2\theta\,d\phi^2,
\end{eqnarray}
where $\Delta^*=r^2-2rf(u)+a^2+e^2$. The Weyl scalar of this
metric becomes
\begin{eqnarray}
\psi_2={1\over\overline R\,\overline
R\,R^2}\Big\{e^2-R\,f(u)\Big\}.
\end{eqnarray}
and $\psi_3$, $\psi_4$ are unaffected as (4.4). From (3.4) we know
that the remaining metric (4.11) is the rotating charged Vaidya
null radiating black hole with $f(u)>a^2+e^2$. The surface gravity
of the horizon at $r=r_+= f(u)+\surd{\{f(u)^2-a^2-e^2\}}$, is 
\begin{eqnarray*}
{\cal K}=-\frac{1}{r_+R^2}\Big[r_+\sqrt{\{f(u)^2-a^2
-e^2\}}+\frac{e^2}{2}\Big]. 
\end{eqnarray*}
The Hawking's temperature on the horizon is $T_{\rm H}=
({\cal K}/2\pi)$. The entropy and angular velocity of the horizon are
found
\begin{eqnarray*}
&&{\cal S}=2\pi\,f(u)\Big\{f(u)+\sqrt{f(u)^2
-(a^2+e^2)}\,\Big\}-\frac{e^2}{4}, \cr
&&\Omega_{\rm H}=\frac{a\{2rf(u)-e^2(u)\}}{(r^2+a^2)^2}\Big|_{r=r_{+}}.
\end{eqnarray*}
Thus, we may regard this left out remnant of the Hawking 
evaporation as the rotating charged Vaidya black hole. On the 
other hand, the metric (4.11)
may be interpreted as the presence of Vaidya mass function $f(u)$
can avoid the formation of an `instantaneous' naked singularity
with zero mass. The formation of `instantaneous' naked singularity
with zero mass - {\it a naked singularity that exists for an
instant and then continues its electrical radiation to create
negative mass}, in {\sl non-embedded} Reissner-Nordstrom and
Kerr-Newman, black holes is unavoidable during Hawking's
evaporation process, as shown in [1]. That is, if we set the mass
function $f(u)=0$, the metric (4.11) would certainly represent an
`instantaneous' naked singularity with zero mass, and at that
stage gravity of the surface would depend only on electric charge,
{\it i.e.} $\psi_2=(e^2/{\overline R\,\overline R\,R^2)}$, and not
on the mass of black hole. However, the Maxwell scalar $\phi_1$ is
unaffected. Thus, from (4.11) with $f(u)\neq 0$ it seems natural
to refer to the {\it rotating} charged Vaidya null radiating black
hole as an `instantaneous' black hole - {\it a black hole that
exists for an instant and then continues its electrical
radiation}, during the Hawking's evaporation process of
Kerr-Newman-Vaidya black hole.

   The time taken between two consecutive radiations is supposed
to be so short that one may not physically realize how quickly
radiations take place. Immediately  after the exhaustion of the
Kerr-New-\\man mass, if one continues the remaining solution
(4.11) to radiate electrically with $e(r)$, there may be a
formation of new mass $m^*_1$ (say). If this electrical radiation
process continues forever, the new mass will increase gradually as
\begin{equation}
{\cal M}^*=m^*_1 + m^*_2 + m^*_3 + m^*_4 + . . .    .
\end{equation}
However, it appears that this new mass will never decrease.
Then, the spacetime metric will take the following form
\begin{eqnarray}
d s^2&=&[1+R^{-2}\{2r({\cal M}^*-f(u))+e^2\}]\,du^2 \cr
&&+2du\,dr+2aR^{-2}\{2r(f(u)-{\cal M}^*)\cr &&-e^2\}\,{\rm
sin}^2\theta\,du\,d\phi-2a\,{\rm sin}^2\theta\,dr\,d\phi\cr &&
-R^2d\theta^2 -\{(r^2+a^2)^2 \cr &&-\Delta^*a^2\,{\rm
sin}^2\theta\}\,R^{-2}{\rm sin}^2\theta\,d\phi^2,
\end{eqnarray}
where $\Delta^*=r^2-2r\{f(u)-{\cal M}^*\}+a^2+e^2$. This metric
will describe a black hole if $f(u)-{\cal M}^*>a^2+e^2$, that is,
when $f(u)>{\cal M}^*>a^2+e^2$. Thus, we have shown the changes in
the total mass of Kerr-Newman-Vaidya black hole in classical
spacetime metrics without effecting the Maxwell scalar and the
Vaidya mass function, for every electrical radiation during the
primordial Hawking evaporation process. We have also observed
that, when $f(u)>{\cal M}^*$, the presence of Vaidya mass $f(u)$
in (4.14) can prevent the direct formation of negative mass naked
singularity. Otherwise, when $f(u)<{\cal M}^*$, this metric may
describe a `non-stationary' negative mass naked singularity, which
is different from the `stationary' one discussed in [1]. The
metric (4.14) can be written in Kerr-Schild ansatze as
\begin{eqnarray*}
 g_{ab}^{\rm NMV}=g_{ab}^{\rm V}+2Q(r,\theta)\ell_a\ell_b,
\end{eqnarray*}
where $Q(r,\theta) =(r{\cal M}^*+e^2/2)R^{-2}$, and
\begin{eqnarray*}
g_{ab}^{\rm NMV}=g_{ab}^{\rm NM}+2Q(u,r,\theta)\ell_a\ell_b,
\end{eqnarray*}
with $Q(u,r,\theta) =-rf(u)R^{-2}$. These Kerr-Schild forms show
that the  metric (4.27) is a solution of Einstein's field
equations. Here the metric tensor $g_{ab}^{\rm V}$ is rotating
Vaidya null radiating metric, and $g_{ab}^{\rm NM}$ is the metric
describing the negative mass naked singularity.

\vspace{0.15in}

{\it (ii) Variable-charged Kerr-Newman-Vaidya-\\de Sitter black
hole}

\vspace{0.15in}

Hawking radiation of black hole is due to the electrical radiation
described by the energy-momentum tensor of electromagnetic field
with variable charge $e(r)$. In order to incorporate the Hawking
radiation in Kerr-Newman-Vaidya-de Sitter black hole, we consider
the charge $e$ of the electromagnetic field to be a function of
radial coordinate. Accordingly, the metric (3.34) embedded into
the de Sitter space will take the form (by changing the mass
symbol $m$ to $M$)
\begin{eqnarray}
ds^2&=&\Big[1-R^{-2}\Big\{2r\Big(M+f(u)\Big)+{\Lambda^*\,r^4\over
3} \cr &&-e^2(r)\Big\}\Big]\,du^2 +2du\,dr +2aR^{-2}\Big\{2r\Big(M
\cr &&+f(u)\Big)+{\Lambda^*\,r^4\over 3} -e^2(r)\Big\}\,{\rm
sin}^2 \theta\,du\,d\phi \cr &&-2a\,{\rm sin}^2\theta\,dr\,d\phi
-R^2d\theta^2 -\Big\{(r^2+a^2)^2 \cr &&-\Delta^*a^2\,{\rm
sin}^2\theta\Big\}\,R^{-2}{\rm sin}^2\theta\,d\phi^2,
\end{eqnarray}
where $\Delta^*=r^2-2r\{M+f(u)\}-\Lambda^*\,r^4/3+a^2+e^2(r)$ and
$\Lambda^*$ represents the cosmological constant of the de Sitter
space. Now we can obtain the Ricci scalars and the Weyl scalars
for this metric (4.15) from (3.6) and (3.8). However, the
expression of Ricci scalars $\phi_{11}$ and $\Lambda$ are purely
matter dependent -- the cosmological constant $\Lambda^*$ and  the
electric charge $e(r)$ with its derivatives. As the cosmological
object and the electromagnetic field are two different matter
fields of different physical properties, it is, without loss of
generality, possible to have a decomposition of $\phi_{11}$ and
$\Lambda$ in terms of the cosmological constant $\Lambda^*$ and
the electromagnetic field with charge $e$ such that
$\phi_{11}$=$\phi^{(\rm C)}_{11} + \phi^{(\rm E)}_{11}$ and
$\Lambda$=$\Lambda^{(\rm C)}+\Lambda^{(\rm E)}$ with
\begin{eqnarray}
\phi_{11}^{(\rm C)}&=&-{1\over {2\,R^2\,R^2}}\,
{\Lambda^*r^2}\,a^2\,{\rm cos}^2\theta, \\\
\phi_{11}^{(\rm E)}&=&{1\over
{2\,R^2\,R^2}}\,\Big\{e^2(r)-2r\,e(r)e(r)_{,r}\Big\} \cr &&+
{1\over
4R^2}\,\Big\{e^2(r)_{,r}+e(r)\,e(r)_{,rr}\Big\}, \\\
\Lambda^{(\rm C)}& =&{\Lambda^*r^2\over 6\,R^2}, \\\
\Lambda^{(\rm E)} &=&{-1\over 12\,R^2}\,\Big\{e^2(r)_{,r}
+e(r)\,e(r)_{,rr}\Big\},
\end{eqnarray}
and the non-vanishing Weyl scalars are given by
\begin{eqnarray}
\psi_2&=&{1\over\overline R\,\overline
R\,R^2}\Big[e^2(r)-R\{M+f(u)\} \cr &&+{1\over 6}\,\overline
R\,\overline R\,\Big\{e^2(r)_{,r} +e(r)\,e(r)_{,rr}\Big\} \cr
&&-\overline
R\,e(r)\,e(r)_{,r}+{\Lambda^*r^2\over3}a^2\,{\rm cos}^2\theta\Big], \\
\psi_3&=&{-i\,a\,{\rm sin}\theta\over 2\surd 2\overline
R\,\overline R\,R^2}\Big\{(4\,r+\overline R)f(u)_{,u}\Big\}, \cr
\psi_4 &=&{{a^2r\,{\rm sin}^2\theta}\over 2\overline R\,\overline
R\,R^2\,R^2}\,\Big\{R^2f(u)_{,uu}-2rf(u)_{,u}\Big\}. \nonumber
\end{eqnarray}
The decomposition of $\phi_{11}$ and $\Lambda$ is followed from
(3.36) for the two energy momentum tensors $T^{(\rm E)}_{ab}$ and
$T^{(\rm C)}_{ab}$ admitted by Kerr-Newman-Vaidya-de Sitter
solution. Now the vanishing Ricci scalar $\Lambda^{(\rm E)}$
(4.19) for electromagnetic field will give
\begin{equation}
e^2(r)=2rm_1 + C
\end{equation}
where $m_1$ and $C$ are real constants. The substitution of this
value in (4.17) and (4.20) with the identification of the constant
$C\equiv e^2$ yields that
\begin{eqnarray}
&\phi^{(\rm E)}_{11}=&{e^2\over {2\,R^2\,R^2}}, \\\
&\psi_2=&{1\over\overline R\,\overline R\,R^2}\Big[e^2-R\{(M-m_1)
\cr &&+f(u)\}+{\Lambda^*r^2\over3}a^2\,{\rm cos}^2\theta\Big].
\end{eqnarray}
We have seen the change in the mass in (4.23) by a quantity $m_1$
(say) at the end of the first electrical radiation. However, due
to the equation (4.22) the Maxwell scalar $\phi_1$  with constant
charge $e$ remains the same as before $\phi_{1}={e/{(\surd
2\,\overline R\,\overline R)}}$ and also the cosmological Ricci
scalar $\Lambda^{(\rm C)}$ unchanged (4.18). Then the total mass
of the Kerr-Newman-Vaidya-de Sitter black hole will take the form
$(M-m_1)+f(u)$, and the line element will be of the form after the
first radiation
\begin{eqnarray}
ds^2&=&\Big[1-R^{-2}\Big\{2r\Big(M-m_1+f(u)\Big)+{\Lambda^*\,r^4\over
3} \cr &&-e^2\Big\}\Big]\,du^2 +2du\,dr +2aR^{-2}\Big\{2r\Big(M
\cr &&-m_1+f(u)\Big)+{\Lambda^*\,r^4\over 3} -e^2\Big\}\,{\rm
sin}^2 \theta\,du\,d\phi \cr &&-2a\,{\rm sin}^2\theta\,dr\,d\phi
-R^2d\theta^2 -\Big\{(r^2+a^2)^2 \cr &&-\Delta^*a^2\,{\rm
sin}^2\theta\Big\}\,R^{-2}{\rm sin}^2\theta\,d\phi^2,
\end{eqnarray}
where $\Delta^*=r^2-2r\{(M-m_1)+f(u)\}-\Lambda^*\,r^4/3+a^2+e^2$.
For the second radiation we consider the charge $e$ to be the
function of $r$ in Einstein's field equations with the mass
$(M-m_1)+f(u)$. Here we again calculate the Ricci scalar of
electromagnetic field $\Lambda^{(\rm E)}$ which has to vanish to
reduce another constant $m_2$ (say), such that after the second
radiation of the black hole (4.24), the total mass will take the
form $M-(m_1-m_2)+f(u)$. However, the Maxwell scalar $\phi_1$, the
cosmological Ricci scalar $\Lambda^*$ and the Vaidya mass $f(u)$
are unaffected by the second electrical radiation. Hence, if we
consider n electrical radiations with the charge $e$ to be
function of $r$, the Einstein's field equations would yield that
the total mass of the black hole will take the form $M-(m_1 + m_2
+ m_3 + m_4 + . . .+ m_n)+f(u)$. Here it is emphasized that there
may be two possibilities that (i) the mass of the Kerr-Newman
black hole radiated away, just leaving the total $M$ equivalent to
the Planck mass and $f(u)$ and $\Lambda^*$ unaffected by the
electrical radiation process; $M$ may not be exactly equal to the
reduced mass $m_1 + m_2 + m_3 + m_4 + . . .+ m_n$, leaving
Planck-size mass; or (ii) the mass $M$ is completely evaporated
with $M=m_1 + m_2 + m_3 + m_4 + . . .+ m_n$ just leaving the
Vaidya mass function $f(u)$ behind, embedded into the de Sitter
cosmological space and the electrical charge $e$ of Kerr-Newman
black hole. The remaining remnant will be of the form of spacetime
metric as
\begin{eqnarray}
ds^2&=&\Big[1-R^{-2}\Big\{2rf(u)+{\Lambda^*\,r^4\over 3}
-e^2\Big\}\Big]\,du^2 \cr &&+2du\,dr
+2aR^{-2}\Big\{2r+f(u)+{\Lambda^*\,r^4\over 3}\cr &&
-e^2\Big\}\,{\rm sin}^2\theta\,du\,d\phi -2a\,{\rm
sin}^2\theta\,dr\,d\phi -R^2d\theta^2 \cr &&-\Big\{(r^2+a^2)^2
-\Delta^*a^2\,{\rm sin}^2\theta\Big\}\,R^{-2}{\rm
sin}^2\theta\,d\phi^2, \nonumber \\
\end{eqnarray}
where $\Delta^*=r^2-2rf(u)-\Lambda^*\,r^4/3+a^2+e^2$. The Weyl
scalar describing the curvature of this black hole remnant is
calculated as
\begin{eqnarray}
\psi_2={1\over\overline R\,\overline R\,R^2}\Big[e^2+R\,f(u)
+{\Lambda^*r^2\over3}a^2\,{\rm cos}^2\theta\Big]
\end{eqnarray}
and the other two Ricci scalars $\psi_3$ and $\psi_4$ are the same
as given in (4.4) indicating that the curvature of algebraically
special gravitational field of rotating Vaidya-de Sitter space is
unaffected during the Hawking electrical radiation process of the
mass of Kerr-Newman black hole. The surface gravity of the horizon
at $r=r_{++}$ is
\begin{eqnarray*}
{\cal K}=-\Big[\,\frac{1}{rR^2}\Big\{r\Big(r-f(u)
-\frac{\Lambda^*r^3}{6}\Big)+\frac{e^2}{2}\Big\}\Big]_{r=r_{++}}
\end{eqnarray*}
with the Hawking's temperature $T_{\rm H}= ({\cal K}/2\pi)$ on it. 
The entropy and angular velocity of the horizon are obtained as
\begin{eqnarray*}
&&{\cal S}=\pi\{r^{2}+a^2\}\Big|_{r=r_{++}}, \cr
&&\Omega_{\rm H}=\frac{a\{2rf(u)+(\Lambda^{*}r^4/3)-e^2\}}{(r^2+a^2)^2}\Big|_{r=r_{++}}.
\end{eqnarray*}
Here the value of $r_{++}$ may be obtained from appendix (A5).
If one sets $f(u) = \Lambda^*=0$ in (4.25), one will obtain the
rotating `instantaneous' naked singularity with zero mass, whose
surface gravity depends only on the electric charge $e$ with the
Weyl scalar $\psi_2={(e^2/{\overline R\,\overline R\,R^2)}}$. The
formation of `instantaneous' naked singularity with zero mass can
be prevented in the case of Kerr-Newman-Vaidya-de Sitter black
hole by the presence of Vaidya mass $f(u)$ and the cosmological
constant $\Lambda^*$ during Hawking's radiation process. Since the
metric (4.25) describes the rotating Vaidya-de Sitter black hole,
it can be referred to the rotating charged Vaidya-de Sitter black
hole as an `instantaneous' black hole; that is, it exists for an
instant and then continues its electrical radiation to create
negative mass, during the Hawking's evaporation process of
Kerr-Newman-Vaidya-de Sitter black hole. However, we find that the
Maxwell scalar $\phi_1$ and the cosmological constant $\Lambda^*$
are unaffected during the radiation process, i.e., the metric
(4.25) still admits the total energy momentum tensor (3.35).

It is emphasized that the time taken between two consecutive
radiations is supposed to be so short that it may not be possible
to realize physically how quickly radiation take place.
Immediately after the exhaustion of the Kerr-Newman mass, if the
remaining solution (4.25) continues to radiate electrically with
the variable charge $e(r)$, the Einstein's field equations with
the vanishing Ricci scalar $\Lambda$ will lead to create a new
mass $m^*_1$ (say). If the electrical radiation process  of black
hole (4.25) continues forever, the new mass might increase
gradually as ${\cal M}^*=m^*_1 + m^*_2 + m^*_3 + m^*_4 + . . .$ .
This will lead the classical spacetime metric with this mass
\begin{eqnarray}
d s^2&=&\Big[1+R^{-2}\Big\{2r\Big({\cal M}^*-f(u)\Big) \cr
&&-{\Lambda^*\,r^4\over 3}+e^2\Big\}\Big]\,du^2+2du\,dr\cr
&&+2aR^{-2}\Big\{2r\Big(f(u)-{\cal M}^*\Big)+{\Lambda^*\,r^4\over
3} \cr &&-e^2\Big\}\,{\rm sin}^2\theta\,du\,d\phi -2a\,{\rm
sin}^2\theta\,dr\,d\phi \cr &&-R^2d\theta^2 -\Big\{(r^2+a^2)^2 \cr
&&-\Delta^*a^2\,{\rm sin}^2\theta\Big\}\,R^{-2}{\rm
sin}^2\theta\,d\phi^2,
\end{eqnarray}
where $\Delta^*=r^2-2r\{f(u)-{\cal M}^*\}-{\Lambda^*r^4/
3}+a^2+e^2$. Thus, we have seen the changes in the total mass of
Kerr-Newman-Vaidya-de Sitter black hole in classical spacetime
metrics without effecting the Maxwell scalar, the Vaidya mass
function and the cosmological constant, during the Hawking
evaporation process of electrically radiating black hole. The
metric (4.27) can be expressed in Kerr-Schild ansatz as
\begin{eqnarray*}
 g_{ab}^{\rm NMVdS}=g_{ab}^{\rm
VdS}+2Q(r,\theta)\ell_a\ell_b,
\end{eqnarray*}
where $Q(r,\theta) =(r{\cal M}^*+e^2/2)R^{-2}$, and
\begin{eqnarray*}
g_{ab}^{\rm NMVdS}=g_{ab}^{\rm NM}+2Q(u,r,\theta)\ell_a\ell_b,
\end{eqnarray*}
with $Q(u,r,\theta) =-\{rf(u)+\Lambda^*r^4/6\}R^{-2}$. These
Kerr-Schild forms show that the  metric (4.27) is a solution of
Einstein's field equations. Here the metric tensors $g_{ab}^{\rm
VdS}$ and $g_{ab}^{\rm NM}$ are rotating Vaidya-de Sitter metric
and the negative mass naked singularity metric respectively.

\vspace*{0.25in}

{\it (iii) Variable-charged Kerr-Newman-monopole \\
black hole}

\vspace*{0.25in}

Here we shall study the Hawking radiation of variable charged
Kerr-Newman-monopole black hole derived above, when the electric
charge $e$ is considered to be a function of radial coordinate $r$
in the field equations. The line element (3.45) with $e(r)$ takes
the form:
\begin{eqnarray}
ds^2&=&[1-R^{-2}\{2rM+b\,r^2- e^2(r)\}]\,du^2 \cr
&&+2du\,dr+2aR^{-2}\{2rM+b\,r^2 \cr &&-e^2(r)\}\,{\rm sin}^2
\theta\,du\,d\phi -2a\,{\rm sin}^2\theta\,dr\,d\phi \cr
&&-R^2d\theta^2 -\{(r^2+a^2)^2 \cr &&-\Delta^*a^2\,{\rm
sin}^2\theta\}\,R^{-2}{\rm sin}^2\theta\,d\phi^2,
\end{eqnarray}
where $\Delta^*=r^2(1-b)-2rM+a^2+e^2(r)$ and the monopole constant
$b$ lies in $0<b<1$. This metric will reduce to Kerr-Newman
solution when $e$ becomes constant initially and $b=0$, Then, the
Einstein-Maxwell field equations for the metric (4.28) with $e(r)$
can be solved to obtain the following quantities:
\begin{eqnarray}
\phi_{11}&=&{1\over
{2\,R^2\,R^2}}\,\Big\{e^2(r)-2r\,e(r)e(r)_{,r}\Big\} \cr
&&+{1\over 4R^2}\,\Big\{e^2(r)_{,r}+e(r)\,e(r)_{,rr}\Big\} \cr &&
+{1\over {4\,R^2\,R^2}}\,b\Big(r^2-a^2{\rm cos}^2\theta\Big), \\\
\Lambda &=& -\frac{1}{
12\,R^2}\,\Big\{e^2(r)_{,r}+e(r)\,e(r)_{,rr}\Big\}
\cr && +\frac{b}{12\,R^2}, \\\
 \psi_2&=&{1\over\overline R\,\overline R\,R^2}\Big[e^2(r)-RM
-\overline R\,e(r)\,e(r)_{,r}\cr &&+{1\over 6}\,\overline
R\,\overline R\,\Big\{e^2(r)_{,r} +e(r)\,e(r)_{,rr}\Big\} \cr
&&-\frac{b}{6}\Big\{RR+2ia\,r\,{\rm cos}\,\theta\Big\}\Big].
\end{eqnarray}
According to the total stress-energy momentum tensor (3.46) we
shall, without loss of generality, have the following
decompositions
\begin{eqnarray}
\phi^{(\rm E)}_{11}&=&{1\over
{2\,R^2\,R^2}}\,\Big\{e^2(r)-2r\,e(r)e(r)_{,r}\Big\} \cr
&&+{1\over 4R^2}\,\Big\{e^2(r)_{,r}+e(r)\,e(r)_{,rr}\Big\}, \\\
\phi^{(\rm
m)}_{11}&=&{1\over {4\,R^2\,R^2}}\,b\Big(r^2-a^2{\rm cos}^2\theta\Big),\\\
\Lambda^{(\rm E)} &=& \frac{-1}{
12\,R^2}\,\Big\{e^2(r)_{,r}+e(r)\,e(r)_{,rr}\Big\},\\\
\Lambda^{(\rm m)}&=&\frac{b}{ 12\,R^2}.
\end{eqnarray}
For electromagnetic field, the Ricci scalar $\Lambda^{(\rm E)}$
given in (4.34) must vanish. This yields
\begin{equation}
e^2(r)=2rm_1 + C
\end{equation}
with $m_1$ and $C$ real constant of integration. Then we
substitute this results in (4.32) to get
\begin{eqnarray}
\phi^{(\rm E)}_{11}&=&{C\over {2\,R^2\,R^2}}.
\end{eqnarray}
Now replacing $C$ by a real constant $e^2$ in (4.37) we obtain the
Maxwell scalar
\begin{eqnarray}
\phi_{1}={e\over {\surd 2\,\overline R\,\overline R}}
\end{eqnarray}
with the charge $e$. Then using the relation (4.36) in (4.31), we
find the changed Weyl scalar
\begin{eqnarray}
\psi_2={1\over\overline R\,\overline R\,R^2}\Big[e^2-R(M-m_1) \nonumber \\
-\frac{b}{6}\Big\{RR+2ia\,r\,{\rm cos}\,\theta\Big\}\Big].
\end{eqnarray}
which shows the reduction of the mass $M$ by some quantity $m_1$.
Thus, we have the line element with change of mass as
\begin{eqnarray}
ds^2&=&[1-R^{-2}\{2r(M-m_1)+b\,r^2- e^2\}]\,du^2 \cr
&&+2du\,dr+2aR^{-2}\{2r(M-m_1)+b\,r^2 \cr &&-e^2\}\,{\rm sin}^2
\theta\,du\,d\phi -2a\,{\rm sin}^2\theta\,dr\,d\phi \cr
&&-R^2d\theta^2-\{(r^2+a^2)^2 \cr &&-\Delta^*a^2\,{\rm
sin}^2\theta\}\,R^{-2}{\rm sin}^2\theta\,d\phi^2,
\end{eqnarray}
with $\Delta^*=r^2(1-b)-2r(M-m_1)+a^2+e^2$. The introduction of
the constant $m_1$ in the metric (4.40) suggests that the first
electrical radiation of Kerr-Newman-monopole black hole has
reduced the original mass $M$ by a quantity $m_1$. For the next
radiation, we again consider the charge $e$ to be the function of
$r$ with the mass $M-m_1$ in (4.40). Then the Einstein-Maxwell
equations yield to reduce the mass by another constant quantity
$m_2$ (say), i.e., after the second radiation, the mass becomes
$M-(m_1+m_2)$. Thus , if we consider $n$ radiations, every time
considering the charge $e$ to be a function of $r$, the Maxwell
scalar $\phi_1$ will be the same as in (4.38). However the mass
will become $M-(m_1 + m_2 + m_3 + m_4 + . . .+ m_n)$ without
affecting monopole constant $b$. Taking Hawking's radiation of
charged black holes into account, one might expect that the mass
$M$ may be radiated away, just leaving $M-(m_1 + m_2 + m_3 + m_4 +
. . .+ m_n)$ equivalent to the Planck mass. Otherwise, the mass
$M$ may be evaporated completely after continuous radiation, when
$M=(m_1 + m_2 + m_3 +m_4+ . . .+ m_n)$, just leaving the monopole
constant $b$ and the electric charge $e$. Thus, we can show this
situation of the black hole in the form of a classical spacetime
metric as
\begin{eqnarray}
ds^2&=&[1-R^{-2}\{b\,r^2- e^2\}]\,du^2+2du\,dr \cr
&&+2aR^{-2}\{b\,r^2-e^2\}\,{\rm sin}^2 \theta\,du\,d\phi \cr
&&-2a\,{\rm sin}^2\theta\,dr\,d\phi-R^2d\theta^2-\{(r^2+a^2)^2 \cr
&&-\Delta^*a^2\,{\rm sin}^2\theta\}\,R^{-2}{\rm
sin}^2\theta\,d\phi^2,
\end{eqnarray}
where $\Delta^*=r^2(1-b)+a^2+e^2$. At this stage the gravity of
the surface would depend on electric charge and monopole constant,
not on the mass. This metric has the non-zero Weyl scalar
\begin{eqnarray}
\psi_2={1\over\overline R\,\overline R\,R^2}\Big[e^2
-\frac{b}{6}\Big\{RR+2ia\,r\,{\rm cos}\,\theta\Big\}\Big],
\end{eqnarray}
describing the curvature of the remaining remnant. However, the
Maxwell scalar $\phi_1$ is unaffected. The metric describes the
rotating charged monopole black hole with the horizons at
$r_{\pm}=\pm\frac{1}{(1-b)}[\surd \{(b-1)(a^2+e^2)\}]$. As the
electrical radiation has to continue, this black hole will remain
only for an instant. Hence, one can refer to the solution (4.41)
as an `instantaneous' charged black hole with the surface gravity
\begin{eqnarray*}
{\cal K}=-\frac{1}{r_+R^2}\Big[r_+\Big\{r_+
\Big(1-\frac12\,b\Big)\Big\}+\frac{e^2}{2}\Big]
\end{eqnarray*}
and the Hawking's temperature $T_{\rm H}= ({\cal K}/2\pi)$. 
The entropy and angular velocity of the horizon are given as
\begin{eqnarray*}
&&{\cal S}=\pi\{r^{2}+a^2\}\Big|_{r=r_{+}},\cr
&&\Omega_{\rm H}=\frac{a\{br^2-e^2(u)\}}{(r^2+a^2)^2}\Big|_{r=r_{+}}.
\end{eqnarray*}
Immediately,  after the exhaustion of the Kerr-Newman mass, if the
remaining solution (4.41) continues to radiate electrically with
$e(r)$, there will be a formation of new mass $m^*_1$ (say). If
this electrical radiation process continues forever, the new mass
will increase gradually as
\begin{equation}
{\cal M}^*=m^*_1 + m^*_2 + m^*_3 + m^*_4 + . . .
\end{equation}
However, it appears that this new mass will never decrease. Then,
the spacetime metric will take the following form
\begin{eqnarray}
ds^2&=&[1+R^{-2}\{2r{\cal M}^*-b\,r^2+ e^2\}]du^2+2du\,dr \cr
&&+2aR^{-2}\{b\,r^2-2r{\cal M}^*-e^2\}\,{\rm sin}^2
\theta\,du\,d\phi \cr &&-2a\,{\rm
sin}^2\theta\,dr\,d\phi-R^2d\theta^2-\{(r^2+a^2)^2 \cr &&
-\Delta^*a^2\,{\rm sin}^2\theta\}\,R^{-2}{\rm
sin}^2\theta\,d\phi^2,
\end{eqnarray}
where $\Delta^*=r^2(1-b)+2r{\cal M}^*+a^2+e^2$. This metric has
the Weyl scalar
\begin{eqnarray}
\psi_2={1\over\overline R\,\overline RR^2}\Big[R{\cal M}^*+e^2
-\frac{b}{6}\{RR+2iar{\rm cos}\theta\}\Big] \nonumber \\
\end{eqnarray}
describing the gravity of the surface with negative sign. This
metric (4.44) expresses the negative mass naked singularity
embedded into the rotating mono-\\pole solution i.e., the metric
can be expressed in Kerr-Schild ansatz based on different
backgrounds as
\begin{equation}
g_{ab}^{\rm NMm}=g_{ab}^{\rm m}+2Q(r,\theta)\ell_a\ell_b
\end{equation}
where $Q(r,\theta) =(r{\cal M}+e^2/2)R^{-2}$, and
\begin{equation}
g_{ab}^{\rm KNm}=g_{ab}^{\rm NM}+2Q(r,\theta)\ell_a\ell_b
\end{equation}
with $Q(r,\theta) =-(b\,r/4)R^{-2}$. Here $g_{ab}^{\rm m}$ is the
rotating monopole solution and $g_{ab}^{\rm NM}$ represents the
negative mass naked singularity metric tensor. The vector $\ell_a$
is a geodesic, shear free, expanding as well as non-zero twist
null vector which is one of the repeated principal null vectors of
both $g_{ab}^{\rm m}$ as well as $g_{ab}^{\rm NM}$. These two
Kerr-Schild forms indicate that the metric (4.44) is a solution of
Einstein's field equations.

\begin{center}
{\bf 5. CONCLUSION}
\end{center}
\setcounter{equation}{0}
\renewcommand{\theequation}{5.\arabic{equation}}

In this paper, we have presented NP quantities for a rotating
spherically symmetric metric with three variables in an appendix.
We find that the general expressions in NP quantities can be used
to discuss the general properties of the spacetimes. For example,
the metric (2.3) with three variables, in general possesses a
geodesic, shear free, rotating and expanding null vector $\ell_a$
as shown in (2.8) and (2.9). The non-vanishing $\psi_2$, $\psi_3$,
$\psi_4$, presented in appendix (A2) suggests that the spacetime
metric (2.3) is algebraically special in the Petrov
classification. With the help of these NP quantities, we have
first given a class of rotating solutions like, rotating
Vaidya-Bonnor, rotating Vaidya, Kerr-Newman and rotating de
Sitter. Then, with the help of Wang-Wu functions, we come to the
unpublished examples of rotating metrics that we combined them
with other rotating solutions in order to generate new embedded
rotating solutions like Kerr-Newman-Vaidya, Kerr-Newman-Vaidya-de
Sitter, Kerr-Newman-monopole, and studied the gravitational
structure of the solutions by observing the nature of the energy
momentum tensors of respective spacetime metrics. The embedded
rotating solutions have also been expressed in terms of
Kerr-Schild ansatze in order to indicate them as solutions of
Einstein's field equations. These ansatze show the extensions of
those of Glass and Krisch [7] and Xanthopoulos [8].

The remarkable feature of the analysis  of rotating solutions in
this paper is that all the rotating solutions, stationary Petrov
type $D$ and non-stationary algebraically special possess the same
null vector $\ell_a$, which is geodesic, shear free, expanding as
well as non-zero twist. From the studies of the rotating solutions
we find that some solutions after making rotation have disturbed
their gravitational structures. For example, the rotating monopole
solution (3.45) with $m=e=0$ possesses the energy momentum tensor
with the monopole pressure $p$, where the monopole charge $b$
couples with the rotating parameter $a$. Similarly, the rotating
de Sitter solution (3.25) becomes Petrov type $D$ spacetime
metric, where the rotating parameter $a$ is coupled with the
cosmological constant. After making rotation in (3.4), the Vaidya
metric with $e(u)=0$ becomes algebraically special in Petrov
classification of spacetime metric with  a null vector $\ell_a$
which is geodesic, shear free, expanding and non-zero twist. The
Wang-Wu functions in the rotating metric (3.10) play a great role
in the derivation of the rotating embedded solutions discussed
here. The method adopted here with Wang-Wu functions might be
another possible version for obtaining {\sl non-stationary}
rotating black hole solutions with visible energy momentum tensors
describing the interaction of different matter fields with
well-defined physical properties like Guth's modification of
$T_{ab}$ (3.39) etc. It is believed that such interactions of
different matter fields as in (3.16), (3.35), (3.39) and (3.46)
have not seen published before. We have also found the direct 
involvement of the rotation parameter $a$ in each expression
of the surface gravity and the angular velocity, which shows 
the important of the study of rotating, embedded and non-embedded,
black holes in order to understand the nature of different black 
holes located in the universe.

In section 4, we find that the changes in the masses of embedded
black holes take place due to the vanishing of Ricci scalar of
electromagnetic fields with the charge $e(r)$. It is also shown
that the Hawking's radiation can be expressed in classical
spacetime metrics, by considering the charge $e$ to be the
function of the radial coordinate $r$ of Kerr-Newman-Vaidya,
Kerr-Newman-Vaidya-de Sitter and Kerr-Newman-monopole black holes.
That is, every electrical radiation produces a change in the mass
of the charged objects. These changes in the mass of black holes
embedded into Vaidya, Vaidya-de Sitter and monopole spaces, after
every electrical radiation, describe the relativistic aspect of
Hawking's evaporation of masses of black holes in the classical
spacetime metrics.  Thus, we find that the black hole evaporation
process is due to the electrical radiation of the variable charge
$e(r)$ in the energy momentum tensor describing the change in the
mass in classical spacetime metrics which is in agreement with
Boulware's suggestion [3]. The Hawking's evaporation of masses and
the creation of embedded negative mass naked singularities are
also due to the continuous electrical radiation. The formation of
embedded naked singularity of negative mass is also Hawking's
suggestion [2] mentioned in the introduction above. This suggests
that, if one accepts the continuous electrical radiation to lead
the complete evaporation of the original mass of black holes, then
the same radiation will also lead to the creation of new mass to
form negative mass naked singularities. This clearly indicates
that an electrically radiating  embedded black hole will not
disappear completely, which is against the suggestion made in
[2,3,5]. It is noted that we observe the different results from
the studies of {\sl embedded} and {\sl non-embedded} black holes.
In the embedded cases here above, the presence of Vaidya mass in
(4.11), the Vaidya mass and the cosmological constant in (4.25)
and the monopole charge in (4.41) completely prevent the
disappearance of embedded radiating black holes during the
radiation process and thereby, the formation of `instantaneous'
charged black holes. In {\sl non-embedded} cases in [1], the
disappearance of a black hole during radiation process is
unavoidable, however occurs for an instant with the formation of
`instantaneous' naked singularity with zero mass, before
continuing its next radiation. It is also noted that the
`instantaneous' black holes (4.11), (4.25) and (4.41) admit the
total energy momentum tensors (3.16), (3.35) and (3.46)
respectively as these tensors are not affected by the Hawking's
radiation.

It appears that (i) the changes in the mass of black holes, (ii)
the formation of `instantaneous' naked singularities  with zero
mass and (iii) the creation of `negative mass naked singularities'
in {\sl non-embedded} Reissner-Nordstrom as well as Kerr-Newman
black holes [1] are presumably the correct formulation in
classical spacetime metrics of the three possibilities of black
hole evaporation suggested by Hawking and Israel [34]. However,
the creation of `negative mass naked singularities' may be a
violation of Penrose's cosmic censorship hypothesis [19]. It is
found that (i) the changes in the masses of embedded black holes,
(ii) the formation of `instantaneous' charged black holes (4.11),
(4.25) and (4.41), and (iii) the creation of embedded `negative
mass naked singularities' in Kerr-Newman-Vaidya,
Kerr-Newman-Vaidya-de Sitter, Kerr-Newman-monopole black holes
might presumably be the mathematical formulations in classical
spacetime metrics of the three possibilities of black hole
evaporation [34]. All embedded black holes discussed here can be
expressed in Kerr-Schild ansatze, accordingly their consequent
negative mass naked singularities are also expressible in
Kerr-Schild forms showing them as solutions of Einstein's field
equations. It is also observed that once a charged black hole is
embedded into some spaces, it will continue to embed forever
through out its Hawking evaporation process. For example,
Kerr-Newman black hole is embedded into the rotating Vaidya null
radiating universe, it continues to embed as `instantaneous'
charged black hole in (4.11) and embedded negative mass naked
singularity as in (4.14). There Hawking's radiation does not
affect the Vaidya mass through out the evaporation process of
Kerr-Newman mass. Similarly, in the cases of Kerr-Newman-Vaidya-de
Sitter as well as Kerr-Newman-monopole black holes we find that
the Vaidya mass, the cosmological constant and the monopole charge
remain unaffected. This means that the embedded negative mass
naked singularities (4.14), (4.27) and (4.44) possess the total
energy momentum tensors (3.16), (3.35) and (3.46) respectively, as
the Kerr-Newman mass does not involved in these tensors, and the
change in the mass due to continuous radiation does not affect
them. Thus, it may be concluded that once a black hole is embedded
into some universes, it will continue to embed forever without
disturbing the nature of matters present. If one accepts the
Hawking continuous evaporation of charged black holes, the loss of
mass and creation of new mass are the process of the continuous
radiation. So, it may also be concluded that once electrical
radiation starts, it will continue to radiate forever describing
the various stages of the life of radiating black holes.

Also, we find from the above that the change in the mass of black
holes, {\sl embedded} or {\sl non-embedded}, takes place due to
the Maxwell scalar ${\phi_1}$, remaining unchanged in the field
equations during continuous radiation. So, if the Maxwell scalar
${\phi_1}$ is absent from the space-time geometry, there will be
no radiation, and consequently, there will be no change in the
mass of the black hole. Therefore, we cannot, theoretically,
expect to observe  such {\sl relativistic change} in the mass of
uncharged Schwarzschild as well as Kerr black holes. This suggests
that these uncharged Schwarzschild as well as Kerr black holes
will forever remain the same without changing their life styles.
Therefore, as far as Hawking's radiation effect is concerned, they
may be referred to as {\sl relativistic death black holes}.

From the study of Hawking's radiation above, it is also
found that, as far as the embedded black holes are concerned,
the Kerr-Newman black hole has relations with other
rotating black holes, like the charged Vaidya black hole (4.11),
the charged Vaidya-de Sitter (4.25) and the rotating charged monopole 
(4.41). There the later ones are `instantaneous' black holes of the 
respective embedded ones. It is also noted
that the rotating charged de Sitter, when $f(u)=0$ in (4.25), may
be regarded as an instantaneous cosmological black hole of
Kerr-Newman-de Sitter. It is observed that the classical spacetime
metrics discussed above would describe the possible life style of
radiating embedded black holes at different stages during their
continuous radiation. These {\sl embedded} classical spacetimes metrics
describing the changing life style of black holes are different from the
{\sl non-embedded} ones studied in [1] in various respects shown
above. Here the study of these embedded solutions suggests the
possibility that in an early universe there might be some black
holes, which might have embedded into some other spaces possessing
different matter fields with well-defined physical properties.

\begin{center}
ACKNOWLEDGEMENT
\end{center}
The author acknowledges his appreciation for hospitality
(including Library and Computer) received from Inter University
Centre for Astronomy and Astrophysics (IUCAA), Pune during his
visits.

\section*{Appendix}
\setcounter{equation}{0}
\renewcommand{\theequation}{A\arabic{equation}}

Here we shall present the NP quantities for the metric (2.3),
calculated from the Cartan's first and second equations of structure
developed by McIntosh and Hickman [11] in Newman-Penrose formalism [10].\\

Newman-Penrose spin coefficients:
\begin{eqnarray}
&&\kappa=\sigma=\lambda=\epsilon=0,\cr &&\rho=-{1\over\overline
R},\;\;\;\; \mu=-{H(u,r,\theta)\over{2\overline R}},\cr
&&\alpha={(2ai-R\,{\rm cos}\,\theta)\over{2\surd 2\overline
R\,\overline R\,{\rm sin}\,\theta}},\;\;\; \beta={{\rm
cot}\,\theta\over2\surd 2R}, \cr &&\pi={i\,a\,{\rm
sin}\,\theta\over{\surd 2\overline R\,\overline R}},\;\;\;
\tau=-{i\,a\,{\rm sin}\,\theta\over{\surd 2R^2}}, \\\
&&\gamma={1\over 4}\Big\{H_{,r}-{2ia\,{\rm cos}\,\theta\,H\over
R^2}\Big\},\;\;  \cr && \nu ={1\over{2\surd 2\overline
R}}\Big\{H_{,\theta} - ia{\rm sin}\theta H_{,u} - {2a^2H{\rm
cos}\theta{\rm sin}\theta\over R^2}\Big\}, \nonumber
\end{eqnarray}
where $\Delta^*=r^2-2rM(u,r,\theta)+a^2+e^2(u,r,\theta)$ \\ and
the function $H(u,r,\theta)$ is given in (2.7).

The Weyl scalars:
\begin{eqnarray}
\psi_0&=&\psi_1=0, \cr \psi_2 &=&{1\over{\overline R\,\overline
R\,R^2}}\left\{(-RM+e^2) +\overline R\,(rM_{,r} -ee_{,r})\right\}
\cr  &&+{1\over
6R^2}\Big(-2M_{,r}-rM_{,rr}+e^2_{,r}+ee_{,rr}\Big), \cr \psi_3
&=&{-1\over{2\surd 2\overline R\overline R\,R^2}}\,
\Big[4\Big\{i\,a\,{\rm sin}\,\theta\,(rM_{,u} \cr
&&-ee_{,u})-(rM_{,\theta}-ee_{,\theta})\Big\}  \cr && + {\overline
R}\,\Big\{i\,a\,{\rm sin}\,\theta(r\,M_{,u}-e\,e_{,u})_{,r} \cr
 &&-(r\,M_{,\theta}-e\,e_{,\theta})_{,r}\Big\}\Big],  \\
\psi_4 &=&{\{a\,i(1+2{\rm sin}^2\theta)-r{\rm
cos}\,\theta\}\over{2\overline R\,\overline R\,\overline
R\,R^2\,{\rm sin}\,\theta}}\Big[i\,a\,{\rm sin}\,\theta\,(rM_{,u}
\cr &&-ee_{,u})-(rM_{,\theta}-ee_{,\theta})\Big]- {1\over{\surd
2}\overline R}i\,a\,{\rm sin}\,\theta\,\nu_{,u} \cr
&&+{1\over{\surd 2}\overline R}\,\nu_{,\theta}-{{\surd
2}\over{\overline R\,R^2}}\,a^2\nu\,{\rm sin}\,\theta\,{\rm
cos}\,\theta, \nonumber
\end{eqnarray}
where $\nu$ is given in (A1).

The Ricci scalars:
\begin{eqnarray}
\phi_{00}&=&\phi_{01}=\phi_{10}=\phi_{20}=\phi_{02}=0, \cr
\phi_{11}&=&{1\over{4\,R^2\,R^2}}\,\Big[ 2\,e^2
+4r(r\,M_{,r}-ee_{,r}) \cr
&&+R^2\Big(-2M_{,r}-r\,M_{,r\,r}+e^2_{,r}+ee_{,r\,r}\Big)\Big],\cr
\phi_{12}&=&{1\over{2\surd 2R^2R^2}}\,\Big[ia{\rm sin}\,\theta
\Big\{(RM_{,u}-2ee_{,u}) \cr &&-(rM_{,r}-ee_{,r})_{,u}\overline
R\Big\} \cr &&+\Big\{(RM_{,\theta}-2\,ee_{,\theta})-(
rM_{,r}-ee_{,r})_{,\theta}\overline R\Big\}\Big], \cr
\phi_{21}&=&{-1\over{2\surd 2R^2R^2}}\Big[ia\,{\rm
sin}\,\theta\Big\{(\overline RM_{,u}-2ee_{,u}) \cr
&&-(rM_{,r}-ee_{,r})_{,u}R\Big\} \cr &&+\Big\{(\overline
RM_{,\theta}-2ee_{,\theta})
-(rM_{,r}-ee_{,r})_{,\theta}R\Big\}\Big], \cr
\phi_{22}&=&{1\over{\surd 2}\,R}i\,a\,{\rm sin}\,\theta\,\nu_{,u}
+{1\over{\surd 2}\,R}\,\nu_{,\theta} \cr &&-{{\surd
2}\over{R\,R^2}}\,a^2\nu\,{\rm sin}\,\theta\,{\rm cos}\,\theta \cr
&&-{{(a^2\,{\rm sin}\,\theta+\overline R\,{\rm
cot}\,\theta)}\over{2\overline R\,R^2\,R^2}}\,\Big\{i\,a\,{\rm
sin}\,\theta\,(r\,M_{,u} \cr
&&-e\,e_{,u})\Big\}-{{r^2+a^2}\over{\overline
R\,R^2\,R^2}}\Big(r\,M_{,u}-e\,e_{,u}\Big) \cr &&-{{(a^2\,{\rm
sin}\,\theta-\overline R\,{\rm cot}\,\theta)}\over{2\overline
R\,R^2\,R^2}}{\Big(r\,M_{,\theta}-e\,e_{,\theta}\Big)}, \cr
\Lambda &=&{1\over{12\,R^2}}\,\Big(2M_{,r}+
r\,M_{,r\,r}-e^2_{,r}-ee_{,r\,r}\Big).
\end{eqnarray}

According to Carter [30] and York [35], we shall introduce a
scalar ${\cal K}$ defined by the relation $n^b \nabla_{b}n^a={\cal
K}\,n^a$, where the null vector $n^a$ in (2.5) above is
parameterized by the coordinate $u$, such that $d/du=n^a\nabla_{a}$.
Then this scalar can, in general, be expressed in terms of NP spin
coefficient $\gamma$ (A1) as follows:
\begin{eqnarray}
{\cal K}&=& n^b\,\nabla_{b}\,n^a\ell_a = -(\gamma+\overline\gamma).
\end{eqnarray}
On a horizon, the scalar ${\cal K}$ is called the surface gravity
of a black hole.

The biquadratic equation for the solution (4.24)
\begin{eqnarray*}
\Delta^{*}\equiv r^2-2r\{m+f(u)\}-\Lambda^{*}\,r^4/3+a^2+e^2=0
\end{eqnarray*}
has the following four roots for non-zero cosmological constant $\Lambda^{*}$:
\begin{eqnarray}
\Big[r_{+}\Big]_{\pm}&=&+{\frac{1}{2}}\surd\,{\Gamma} \pm {\frac{1}{2}}
\sqrt\Big[\,\frac{4}{\Lambda^{*}}- 32^{1/3}\chi \cr
&&-\frac{1}{32^{1/3}\Lambda^{*}}\,\Big\{P+\sqrt{P^2-4Q}\Big\}^{1/3}\cr
&&-\frac{12}{\Lambda^{*}\surd\,\Gamma}\,\{m +f(u)\}\Big], 
\end{eqnarray}
\begin{eqnarray}
\Big[r_{-}\Big]_{\pm}&=&-{\frac{1}{2}}\surd\,{\Gamma} \pm {\frac{1}{2}}
\sqrt\Big[\,\frac{4}{\Lambda^{*}}- 32^{1/3}\chi \cr
&&-\frac{1}{32^{1/3}\Lambda^{*}}\,\Big\{P+\sqrt{P^2-4Q}\Big\}^{1/3}\cr
&&+\frac{12}{\Lambda^{*}\surd\,\Gamma}\,\{m +f(u)\}\Big]
\end{eqnarray}
where
\begin{eqnarray*}
&&P=54\Big[18\Lambda^{*}\{m+f(u)\}^2-12\Lambda^{*}(a^2+e^2)-1\Big],\cr
&&Q=\Big\{9-36\Lambda^{*}(a^2+e^2)\Big\}^3, \cr
&&\chi=\frac{1-4\Lambda^{*}(a^2+e^2)}{\Lambda^{*}\Big\{P+\sqrt{P^2-4Q}\Big\}^{1/3}},\cr
&&\Gamma=\frac{2}{\Lambda^{*}}+ 32^{1/3}\chi 
-\frac{32^{-1/3}}{\Lambda^{*}}\Big\{P+\sqrt{P^2-4Q}\Big\}^{1/3}.
\end{eqnarray*}
The calculation of these roots has been carried out by using `Mathematica'.
The area of a horizon of black hole can be calculated as follows [19]:
\begin{eqnarray}
{\cal A} = \int_{0}^{\pi}\int_{0}^{2\pi}\sqrt{g_{\theta\theta}g_{\phi\phi}}
\,d\theta\,d\phi\,\Big|_{\Delta^{*}=0},
\end{eqnarray}
depending on the values of the roots of $\Delta^{*}=0$.
Then the entropy on a horizon of a black hole may be obtained from the relation 
${\cal S}= {\cal A}/4$ [36].

\begin{center}
\line(17,0){210.94}
\end{center}

\end{twocolumn}


\begin{thebibliography}{99}

\bibitem{x}N. Ibohal, Class. Quantum Grav. {\bf 19},
4327 (2002), ({\sl Preprint}) gr-qc/0405019
\bibitem{x}S. W. Hawking, Nature, {\bf 248}, 30 (1974);
Commun. math. Phys. {\bf 43}, 199 (1975);
Phys. Rev. D {\bf 14}, 2460 (1976).
\bibitem{x}D. G. Boulware, Phys. Rev. D {\bf 13}, 2169 (1976).
\bibitem{x} B. Steinmular, A. R. King and J. P. LoSota
Phys. Lett. {\bf 51A}, 191 (1975).
\bibitem{x}F. J. Tipler, C. J. S. Clerke and G. F. R. Ellis
`Singularities and Horizons -- A review article' in {\sl
General Relativity and Gravitation: One Hundred Year After the birth of
Albert Einstein}, Vol 2, edited by A. Held (Plenum Press, New
York, 1980).
\bibitem{x}R. G. Cai, J. Y. Ji and K. S. Soh, Class. Quantum Grav.
{\bf 15}, 2783 (1998).
\bibitem{x}E. N. Glass and J. P. Krisch, Phys. Rev. D {\bf
57}, R5945 (1998); Class. Quantum Grav. {\bf 16}, 1175 (1999).
\bibitem{x}B. C. Xanthopoulos,  J. Math. Phys. {\bf 19},
1607 (1978).
\bibitem{x}A. Wang and Y. Wu, Gen. Rel. Grav. {\bf
31}, 107 (1999)
\bibitem{x}E. T. Newman and R. Penrose, J. Math. Phys.
{\bf 3}, 566 (1962).
\bibitem{x}C. B. G. McIntosh and M. S. Hickman, Gen. Rel.
Grav. {\bf 17}, 111 (1985).
\bibitem{x}D. Xu, Class. Quantum. Grav. {\bf 15}, 153 (1998).
\bibitem{x}W. Bonnor and  P. C. Vaidya,  Gen. Rel.
Grav. {\bf 1}, 127 (1970).
\bibitem{x}E. T. Newman and A. I. Janis, J. Math. Phys.
{\bf 6}, 915 (1965); E. T. Newman, E. Couch, K. Chinnapared,
A. Exton, A. Prakash and R. Torrence, J. Math. Phys.
{\bf 6}, 918 (1965).
\bibitem{x}L. Herrera and J. Jimenez, J Math. Phys.
{\bf 23}, 2339 (1982).
\bibitem{x}S. P. Drake and R. Turolla,  Class Quantum.
Grav. {\bf 14}, 1883 (1997).
\bibitem{x}S. P. Drake and P. Szekeres, Gen. Rel. Grav.
{\bf 32}, 445 (2000).
\bibitem{x}S. Yazadjiev, Gen. Rel. Grav. {\bf 32}, 2345 (2000).
\bibitem{x}S. Chandrasekhar, {\it The Mathematical
Theory of Black Holes} (Clarendon Press, Oxford, 1983).
\bibitem{x}V. Husain, Phys. Rev. D {\bf 53}, R1759 (1996)
\bibitem{x}P. S. Joshi, {\it Global Aspects in
Gravitation and Cosmology} (Clarendon, Oxford, 1993)
\bibitem{x}M. Carmeli and M. Kaye, Ann.Phys (N.Y.),
{\bf 102}, 97 (1977); M. Carmeli, {\it Classical Fields, General Relativity
and Gauge Theory} (New York, Wiley, 1982).
\bibitem{x}L. Herrera and J. Martinez, J. Math. Phys.
{\bf 39}, 3260 (1998).
\bibitem{x}L. Herrera, H. Hernandez, L. A. Nunez and U. Percoco,
Class. Quantum. Grav. {\bf 15}, 187 (1998).
\bibitem{x}J. Jing, and Y. Wang, Int. J. Theo. Phys.
{\bf 35}, 1481 (1996).
\bibitem{x}M. Visser, Phys. Rev. D {\bf 46}, 2445 (1992).
\bibitem{x}G. W. Gibbons and S. W. Hawking, Phys. Rev. D {\bf 15},
2738 (1977).
\vspace*{3.12in}
\bibitem{x}S. W. Hawking and G. F. R. Ellis, {\it The large scale
structure of space-time}, Cambridge University Press, Cambridge,
1973.
\bibitem{x}A. H. Guth, 1981 {\sl Phys. Rev D} {\bf 33} 347 (1981).
\bibitem{x}B Carter, in {\it Black holes} edited by C Dewitt
and B.C. Dewitt (New York, Gordon and Breach Science Publ., 1973)
\bibitem{x}R L Mallett, {\sl Phys. Lett A} {\bf 126} 226 (1988)
\bibitem{x}D Xu, {\sl Class. Quantum Grav.} {\bf 15}, 153 (1998)
\bibitem{x}P. C. Vaidya, {\sl Proc. Indian Acad. Sci.} {\bf
A33} 264 (1951); (Reprinted) {\sl Gen. Rel. Grav.} {\bf 31}, 119
(1999).
\bibitem{x}S. W. Hawking and W. Israel, An Introductory
survey' in {\sl General Relativity: An Einstein Centenary Survey}
edited  by  S. W. Hawking and W. Israel (Cambridge University
Press, 1980).
\bibitem{x}J, W. York, in {\sl Quantum Theory of Gravity: Essays
in honor of the 60th Birthday of Bryce S Dewitt} edited by S. M.
Christensen, Adam Hilger Ltd. Bristol, 1984.
\bibitem{x}S. W. Hawking and S. F. Ross, {\sl Phys. Rev. D} {\bf 52},
5865 (1995).
\end{thebibliography}
\end{document}